\documentclass[aoas,MSNbibl,nameyear,dvips]{arximspdf}
\usepackage{mathbh}
\usepackage{dcolumn}
\usepackage{graphicx}

\doi{10.1214/14-AOAS729} %kopijuoti is PTS
\volume{8}
\issue{2}
\pubyear{2014}
\firstpage{674}
\lastpage{702}

\makeatletter

\newcommand{\rright}{\right}
\newcommand{\lleft}{\left}
\newcommand{\rrvert}{\vert}
\newcommand{\llvert}{\vert}

\newtheorem{corollary}{Corollary}
\newtheorem{theorem}{Theorem}
\newtheorem{prop}{Proposition}

\newproclaim{algo}{Algorithm}

\newcommand{\vsmall}{v}

\newcommand{\Wt}{U_t}
\newcommand{\Vt}{V_t}

\newcommand{\W}{U}
\newcommand{\Wz}{U_0}
\newcommand{\Wn}{U_n}
\newcommand{\Wton}{U_{0:n}}
\newcommand{\Wi}{U_{i}}
\newcommand{\Wim}{U_{i-1}}
\newcommand{\Wip}{U_{i+1}}
\newcommand{\Wtoi}{U_{0:i}}
\newcommand{\Wtoim}{U_{0:i-1}}
\newcommand{\V}{V}
\newcommand{\Vton}{V_{0:n}}
\newcommand{\Vz}{V_0}
\newcommand{\Vi}{V_{i}}
\newcommand{\Vim}{V_{i-1}}
\newcommand{\Vip}{V_{i+1}}
\newcommand{\Vtoi}{V_{0:i}}
\newcommand{\Vtoim}{V_{0:i-1}}
\newcommand{\pDelta}{p_\Delta}
\newcommand{\EDelta}{\mathbb{E}_\Delta}
\newcommand{\E}{\mathbb{E}}
\newcommand{\hthetamm}{\widehat{\theta}_{m-1}}
\newcommand{\hthetam}{\widehat{\theta}_{m}}
\newcommand{\hthetaz}{\widehat{\theta}_{0}}

\newcommand{\Xz}{X_0}
\newcommand{\Xun}{X_1}
\newcommand{\Xnm}{X_{n-1}}
\newcommand{\Xn}{X_n}
\newcommand{\Yun}{Y_1}
\newcommand{\Yn}{Y_n}

\newcommand{\Weighttonk}{W^{(k)}_{0:n}}
\newcommand{\Weightz}{W_0}
\newcommand{\Weighti}{W_i}
\newcommand{\Weight}{W}
\newcommand{\Weightn}{W_n}
\newcommand{\Weightim}{W_{i-1}}

\newcommand{\Wtonk}{U^{(k)}_{0:n}}
\newcommand{\Wtoik}{U^{(k)}_{0:i}}
\newcommand{\Wtoimk}{U^{(k)}_{0:i-1}}
\newcommand{\Wtoiml}{U^{(l)}_{0:i-1}}
\newcommand{\Wik}{U^{(k)}_{i}}
\newcommand{\Wzk}{U^{(k)}_0}

\newcolumntype{d}[1]{D{.}{.}{#1}}

\newcommand{\eqref}[1]{(\ref{#1})}
\newcommand{\var}{\operatorname{var}}
\newcommand{\Cov}{\operatorname{Cov}}

\makeatother

\begin{document}
\begin{frontmatter}

\title{Estimation in the partially observed stochastic Morris--Lecar neuronal
model with particle filter and stochastic approximation methods}
\runtitle{Estimation in neuronal models}

\begin{aug}
\author[A]{\fnms{Susanne}~\snm{Ditlevsen}\corref{}\thanksref{m1,t1}\ead[label=e1]{susanne@math.ku.dk}\ead[label=u1,url]{http://www.math.ku.dk/\textasciitilde susanne}\ead[label=u2,url]{http://dsin.ku.dk}}
\and
\author[B]{\fnms{Adeline}~\snm{Samson}\thanksref{m2,m3,t2}\ead[label=e2]{Adeline.leclercq-Samson@imag.fr}\ead[label=u3,url]{http://adeline.e-samson.org/en/}}
\runauthor{S. Ditlevsen and A. Samson}
\affiliation{University of Copenhagen\thanksmark{m1},  Universit\'e Paris
Descartes\thanksmark{m2} and~Universit\'e~Grenoble-Alpes\thanksmark{m3}}
\thankstext{t1}{Supported in part by grants from the
Danish Council for Independent Research $\mid$ Natural
Sciences.}
\thankstext{t2}{Supported in part by
grants from the University Paris Descartes PCI.}
\address[A]{Department of Mathematical Sciences\\
University of Copenhagen\\
Universitetsparken 5\\
DK-2100 Copenhagen\\
Denmark\\
\printead{e1}\\
\printead{u1}\\
Dynamical Systems Interdisciplinary Network\\
\printead{u2}} %adresu isvedimo komanda gale!
\address[B]{CNRS UMR8145\\
Laboratoire MAP5\\
Universit\'e Paris Descartes\\
45 rue des Saints P\`{e}res\\
75006 Paris\\
France\\
and\\
Laboratoire Jean Kuntzmann UMR\\ CNRS 5224\\
Universit\'e Grenoble-Alpes\\
51 rue des Math\'{e}matiques\\
38041 Grenoble cedex 9\\
France\\
\printead{e2}\\
\printead{u3}}
\end{aug}

% HISTORY:
\received{\smonth{3} \syear{2013}}
\revised{\smonth{10} \syear{2013}}

% ABSTRACT
%
\begin{abstract}
Parameter estimation in multidimensional diffusion models with only one
coordinate observed is highly relevant in many biological applications,
but a statistically difficult problem. In neuroscience, the membrane
potential evolution in single neurons can be measured at high
frequency, but biophysical realistic models have to include the
unobserved dynamics of ion channels. One such model is the
stochastic Morris--Lecar model, defined by a nonlinear two-dimensional
stochastic differential equation. The coordinates are coupled,
that is, the unobserved coordinate is nonautonomous, the model exhibits
oscillations to mimic the spiking behavior, which means it is not of
gradient-type, and the measurement noise from
intracellular recordings is typically
negligible. Therefore, the hidden Markov model framework is degenerate,
and available methods break down.
The main contributions of this paper are an approach to estimate in this
ill-posed situation and nonasymptotic convergence results for
the method. Specifically, we
propose a sequential Monte Carlo particle filter algorithm to impute
the unobserved coordinate, and then estimate parameters maximizing a
pseudo-likelihood through a stochastic version of the
Expectation--Maximization algorithm. It turns out that even the rate scaling
parameter governing the opening and closing of ion channels of the
unobserved coordinate can be reasonably estimated. An
experimental data set of intracellular recordings of the membrane
potential of a spinal motoneuron of a red-eared turtle is analyzed,
and the performance is further evaluated in a
simulation study.
\end{abstract}

% KEYWORDS
\begin{keyword}
\kwd{Sequential Monte Carlo}
\kwd{diffusions}
\kwd{pseudo likelihood}
\kwd{Stochastic Approximation Expectation Maximization}
\kwd{motoneurons}
\kwd{conductance-based neuron models}
\kwd{membrane potential}
\end{keyword}

\end{frontmatter}

%s1 #&#
\section{Introduction}\label{intro}

In neuroscience, it is of major interest to understand the principles
of information processing in the nervous system, and a basic step is to
understand signal processing and transmission in single
neurons. Therefore, there is a growing demand for robust methods to
estimate biophysical relevant parameters from partially observed
detailed models. Statistical inference from experimental data in
biophysically detailed
models of single neurons is difficult. Often these models are compared to
experimental data by hand-tuning to reproduce the
qualitative behaviors observed in experimental data, but without
any formal statistical analysis. It is of particular interest
to estimate conductances, which reflect the synaptic input from the surrounding
network. These can be estimated from intracellular recordings, where the
neuronal membrane potential is recorded at high frequency,
and are typically done
using only subthreshold fluctuations, ignoring the
dynamics during action potentials [\citet
{BergDitlevsen2013,Berg2007,Borg-Graham1998,Monier2008,Pospischil2009,Rudolph2004}].
The
aim of this article is to estimate such biophysical parameters
during the dynamics of spiking from intracellular data.

The Morris--Lecar model [\citet{MorrisLecar1981}] is a simple biophysical
model and a
prototype for a wide variety of neurons. It is a conductance-based
model [\citet{Gerstner2005}], introduced to explain the
dynamics of the barnacle muscle fiber. It is given by two
coupled first order differential equations, the first modeling the
membrane potential evolution and the second the
activation of potassium current. If both current and conductance noise
should be taken into account, the stochastic Morris--Lecar model
arises, where diffusion terms have been added on both coordinates. If
one of these noise sources are zero, a hypoelliptic diffusion arises
leading to singular transition densities and particular statistical
challenges [\citet{Pokern09,SamsonThieullen2012}].
Typically, the membrane potential
will be measured discretely at high frequency, whereas the second
variable cannot be observed. Our goal is to estimate model parameters
from discrete observations of the first coordinate in the nonsingular
case of nonnegligible noise on both coordinates. This includes
estimation of a central rate parameter
characterizing the channel kinetics of the unobserved component, which
we believe has not been done before.

Estimation in these conductance-based models is not
straightforward. Because of the coupling between the coordinates of the
stochastic
differential equation (SDE), the unobserved
coordinate is nonautonomous, and the model does not fit into the
(nondegenerate)
Hidden Markov Model (HMM) framework, as explained in Section~\ref{sec:HMM}. Furthermore, the diffusion is not time reversible and
the likelihood is generally not tractable. Thus, the problem of
inference is complex.
The literature contains various methodologies when all the coordinates
are observed [\citet{Ait-Sahalia2002, Beskos2006, Durham2002,
Jensen2012, Pedersen1995}, \citeauthor{Helle2004} (\citeyear{Helle2004}, \citeyear{Mangabook})] or the hidden state
is Markovian
[\citet{Ionides2011}]. They strongly rely on the Markov property and
are hard to generalize to the non-Markovian case we are studying. In
the non-Markovian case, methods are mainly based on data
augmentation. The idea is that the likelihood can be
approximated given the entire path or a
sufficient partition of it. Therefore, the unobserved coordinates are
treated as missing data and are imputed. Most methods propose to
approximate the transition density by the
Euler--Maruyama scheme and consider a Bayesian point of view to
estimate the posterior distribution of the parameters
[\citet{Elerian2001, Eraker2001},
\citeauthor{Golightly2006}
(\citeyear{Golightly2006},
\citeyear{Golightly2008})]. \citet{Golightly2006}
study a model similar to us but with low frequency data. So
they need to impute data between observations, which is
computationally costly. Furthermore, there exists a strong
dependence between the imputed sample paths and the diffusion
coefficient, and it is not possible to estimate the diffusion parameter with
this kind of approach. An alternative is reparametrization of the
diffusion, but it is limited to scalar diffusions
[\citet{Roberts2001}] or an autonomous hidden coordinate
[\citet{Kalogeropoulos2007}].

In this paper, we propose to estimate the parameters with a maximum
likelihood approach. We approximate the
SDE through an Euler--Maruyama scheme to
obtain a tractable pseudo-likelihood. Then we consider the statistical
model as an incomplete data model and
maximize the pseudo-likelihood through a
stochastic Expectation--Maximization (EM) algorithm,
where the unobserved data are
imputed at each iteration of the algorithm.
We are in the setting of high frequency data so we do not need to
impute data between observations, but our approach could be extended
to that type of data as well.
A similar but different method has been proposed by \citet
{HuysPaninski2006}, where up to $10^4$ parameters are estimated in a
detailed multicompartmental single neuron model. However, only parameters
entering linearly in the loss function are considered, and channel
kinetics are assumed known. It is a quadratic optimization problem
solved by least squares and shown to work well for low noise and high
frequency sampling. When either the discretization step or the noise
increase, a bias is introduced. In \citet{HuysPaninski2009} they
extend the estimation to allow for measurement noise, first smoothing
the data by a particle filter and then maximizing the likelihood
through a Monte Carlo EM algorithm. Because of the measurement
noise, the model fits into the HMM framework and they can use a
standard particle filter. But again, only parameters entering
linearly in the pseudo-likelihood are considered. In particular, all
parameters of the hidden coordinate are assumed known.

Here, we also want to estimate parameters from the hidden coordinate
and we do not consider measurement noise. We propose to impute the
hidden non-Markovian path in the stochastic EM algorithm with a
Sequential Monte Carlo (SMC) algorithm.
Monte Carlo methods for nonlinear filtering are
widely spread, with, among other algorithms, sequential importance
sampling, sequential
importance sampling with resampling (SISR), auxiliary SISR
and stratified resampling [see \citet{cappe05} for a general
presentation]. All SISR algorithms are now called SMC. Most
of them are designed for HMM. In the specific setting of
multidimensional SDEs, \citet{delmoral01} propose a particle filter
for a two-dimensional SDE,
where the second equation is autonomous.
Although the first
coordinate is observed at discrete times, they propose to
simulate it at each iteration of the filter.
%To avoid degeneracy of the weights, they introduce a bounded
%function of the difference between the observed value and the
%simulated particles.
\citet{fearnhead08} generalize this
particle filter to a nonautonomous hidden path but with drift of
gradient type. In the ergodic case this corresponds\vadjust{\goodbreak} to
a time-reversible diffusion. In particular, models exhibiting
oscillations are not covered, which is the case of any realistic
neuronal model.
%
%any proposal distribution to simulate the first coordinate
%at each iteration. The weights then reduce to a Dirac mass of the
%difference between the observed value and the simulated
%particles, which is likely to be almost surely equal to
%zero. Furthermore, they assume the
%drift to be of gradient type. In the ergodic case this corresponds to
%a time reversible diffusion.

These algorithms cannot be directly applied because we are studying a
multidimensional coupled SDE that is not of gradient type. Thus, we
consider the SMC algorithm proposed by \citet{doucet01} for more
general dynamic models than HMM. As we combine this SMC with the
Stochastic Approximation Expectation--Maximization (SAEM) algorithm
which maximizes the pseudo-likelihood based on an Euler--Maruyama
approximation of
the SDE defining the model, we need nonasymptotic convergence
results for the SMC to obtain the convergence of the
SAEM--SMC. Nonasymptotic results for SMC, such as deviation
inequalities, have been proposed in the literature only in the HMM
framework [\citet{DelMoral2000, delmoral01, Douc2011, Kunsch2005}], and
the Markovian structure of the hidden path is a key element in the
proofs. A
major contribution here is that we are able to extend this result to a
SMC for a non-Markovian hidden path.
%Results are obtained with respect to both the filtering distribution
%of the Euler--Maruyama approximate model and the filtering distribution
%of the original SDE.
Then we prove that the
estimator obtained from this
combined SAEM--SMC algorithm converges with probability one to a local
maximum of the pseudo-likelihood. We also prove that the
pseudo-likelihood converges to the true likelihood as the time step
between observations go to zero.

The paper is organized as follows: In Section~\ref{sec:SML} the model is
presented, the noise structure is
motivated, and the pseudo-likelihood arising from the Euler--Maruyama
approximation is found. In Section~\ref{sec:Particle} the filtering
problem is presented, as well as the SMC algorithm and deviation
inequalities. In Section~\ref{sec:estimation} we present the
estimation procedure and the assumptions needed for the convergence
results to hold. In Section~\ref{sec:application} we apply the
method on an experimental data set of intracellular recordings of the
membrane potential of a motoneuron of a turtle, and in Section~\ref{sec:simulation} we conduct a simulation study to
document the performance of the method. Proofs
and technical results can be found in the \hyperref[app]{Appendix}.

%s2 #&#
\section{Stochastic Morris--Lecar model}\label{sec:SML}
%s2.1 #&#
\subsection{Exact diffusion model}

The stochastic Morris--Lecar model including both current and
channel noise is defined as the solution to
%
%e1 #&#
\begin{equation}
\label{eq:SML} \cases{ d\Vt= f(\Vt,\Wt)\,dt +\gamma \,d\tilde{B}_t,
\cr
d\Wt=b(\Vt,\Wt)\,dt + \sigma(\Vt,\Wt)\,dB_t, } %
\end{equation}
where
\begin{eqnarray*}
f(\Vt,\Wt)&=&\frac{1}{C} \bigl(-g_{\mathrm{Ca}}m_\infty(\Vt) (
\Vt -V_{\mathrm{Ca}})-g_{\mathrm{K}}\Wt (\Vt-V_{\mathrm{K}})-g_{\mathrm{L}}(
\Vt-V_{\mathrm{L}})+I \bigr),
\\
b(\Vt,\Wt)&=& \bigl(\alpha(\Vt) (1-\Wt) - \beta(\Vt)\Wt \bigr),
\\
m_\infty(v)&=&\frac{1}{2} \biggl(1+\tanh \biggl(\frac
{v-V_1}{V_2}
\biggr) \biggr),
\\
\alpha(v) &=& \frac{1}{2}\phi\cosh \biggl(\frac{v-V_3}{2V_4} \biggr)
\biggl(1+\tanh \biggl(\frac{v-V_3}{V_4} \biggr) \biggr),
\\
\beta(v) &=& \frac{1}{2}\phi\cosh \biggl(\frac{v-V_3}{2V_4} \biggr)
\biggl(1-\tanh \biggl(\frac{v-V_3}{V_4} \biggr) \biggr),
\end{eqnarray*}
and the initial condition $(\Vz, \Wz)$ is random with density $p(\Vz,
\Wz)$.
Processes $(\tilde{B}_t)_{t\geq t_0}$ and $(B_t)_{t\geq t_0}$ are
independent Brownian motions.
The variable $\Vt$ represents the membrane potential of
the neuron at time $t$, and $\Wt$ represents the normalized
conductance of the K$^+$ current. It varies between 0 and 1,
and can be interpreted as the probability that a K$^+$ ion channel
is open at time $t$. The equation for $f(\cdot)$ describing the
dynamics of
$V_t$ contains four terms,
corresponding to Ca$^{2+}$ current, K$^+$ current, a general leak
current and the input current $I$. The functions
$\alpha(\cdot)$ and $\beta(\cdot)$ model the rates of opening
and closing of the K$^+$ ion channels. The function
$m_{\infty}(\cdot)$ represents the equilibrium value of the
normalized Ca$^{2+}$ conductance for a given value of the membrane
potential. The parameters $V_1, V_2, V_3$ and $V_4$ are scaling
parameters; $g_{\mathrm{Ca}}, g_{\mathrm{K}}$ and $g_{\mathrm{L}}$ are conductances
associated with Ca$^{2+}$, K$^{+}$ and leak
currents; $V_{\mathrm{Ca}}, V_{\mathrm{K}}$ and $V_{\mathrm{L}}$ are reversal
potentials for Ca$^{2+}$, K$^+$ and leak currents; $C$
is the membrane capacitance; $\phi$ is a rate scaling parameter for
the opening and closing of the K$^+$ ion channels; and $I$ is the
input current.

\begin{figure}

\includegraphics{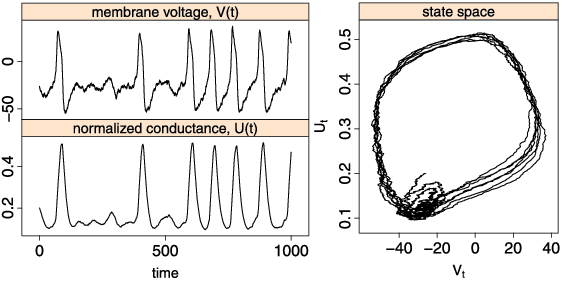}

\caption{Simulated trajectory of the stochastic
Morris--Lecar model: $(\V_t)$ as a function of time (left,
top), $(\W_t)$ as a function of time (left, bottom), and $(\W_t)$
against $(\V_t)$ (right). Parameters are given in Section
\protect\ref{sec:simulation}. Time is measured in ms, voltage in mV,
and the conductance is normalized between 0 and~1.}\label{fig:simu_ML}
\end{figure}

Various noise sources are present in single
neurons, and they act on many different spatial and temporal scales
[\citet{Gerstner2005,Longtin2013}]. A main component arises from the
synaptic bombardment from other neurons in the network, and in the
diffusion limit appears as an additive noise on the current equation.
Parameter $\gamma$ scales this current noise. Conductance fluctuations
caused by random opening and closing of ion channels leads to
multiplicative noise on the conductance equation. Function
$\sigma(\Vt,\Wt)$ models this channel or conductance
noise. We consider the following function that ensures that $\Wt$
stays bounded in the unit interval if $\sigma\leq1$
[\citet{DitlevsenGreenwood2011}]:
$\sigma(\Vt, \Wt) = \sigma\sqrt{2 \frac{\alpha(\Vt) \beta(\Vt
)}{\alpha
(\Vt)+ \beta(\Vt)} \Wt(1-\Wt)}$. A~trajectory of the model is
simulated in Figure~\ref{fig:simu_ML}. The peaks of
$(\Vt)$ correspond to spikes of the neuron.

%f1 #&#

%s2.2 #&#
\subsection{Observations and approximate model}

Data are discrete measurements of $(\Vt)$, while $(\Wt)$ is not
measured. We denote $t_0\leq t_1\leq\cdots\leq t_n$ the discrete
observation times. We denote $\Vi=V_{t_i}$ the observation at time
$t_i$ and $\Vton=(V_{t_0},\ldots,V_{t_n})$ the vector of all the
observed data.
%We focus on the estimation of parameters from observations
%$\Vton$.
Let $\theta\in\Theta\subseteq\mathbb{R}^p$ be the vector of
parameters to be estimated. We consider estimation of
all identifiable parameters of the observed coordinate and the rate
parameter of the unobserved channel dynamics
$\theta=(g_{\mathrm{Ca}}, g_{\mathrm{K}}, g_{\mathrm{L}},
V_{\mathrm{Ca}}, V_{\mathrm{K}}, I, \gamma, \phi)$.
Note that $C$ is
a proportionality factor
of the conductance parameters and thus unidentifiable, as well as the
constant level in $f(\cdot)$ is given by $g_{\mathrm{L}}V_{\mathrm{L}} + I$ and, thus, $V_{\mathrm{L}}$
(or $I$) is unidentifiable. We conjecture that the information about
$\sigma$ in the observed coordinate is close to zero and, thus, in
practice, also $\sigma$ is
unidentifiable from observations of $\Vton$ only, at least for any
finite sample size. This happens because $\sigma$ is mainly shaping the
dynamics of
$\Wt$ between spikes, while the dynamics during spikes resemble
deterministic behavior, and the influence of $\Wt$ on $\Vt$ is only
strong during spikes. This is confirmed in Sections~\ref{sec:application} and \ref{sec:simulation} where misspecification
of $\sigma$ is shown not to deteriorate the estimation of
$\theta$. Finally, we assume the scaling parameters $V_1$--$V_4$ known
because otherwise the model does not belong to an exponential family,
as required by assumption (M1) below. This could be solved by
introducing an extra optimization step in the EM algorithm at
the cost of precision and computer time. It is not pursued further in
this work.

The aim is to estimate $\theta$ by maximum likelihood.
However, this likelihood is intractable, as the transition density of
model (\ref{eq:SML}) is not explicit. %An Euler approximation of the
%model is used.
Let $\Delta$ denote the step size between two observation times, which
for simplicity
we assume do not depend on $i$. The extension to unequally spaced
observation times is straightforward. The Euler--Maruyama approximation
of model (\ref{eq:SML}) leads to a discretized model defined as follows:
%
%e2 #&#
\begin{eqnarray}
\label{eq:discretize_ML_2} \Vip&=& \Vi+\Delta f(\Vi,\Wi)+ \sqrt{\Delta} \gamma\tilde{\eta
}_i,
\nonumber
\\[-8pt]
\\[-8pt]
\Wip&=&\Wi+ \Delta b(\Vi,\Wi) + \sqrt{\Delta}\sigma(\Vi,\Wi )\eta
_i,
\nonumber
%dW_{t}&=&b_\theta(\Vt,\Wt) dt + \sigma_\theta(\Vt,\Wt)dB_t
\end{eqnarray}
where $(\tilde{\eta}_i)$ and $(\eta_i)$ are independent centered
Gaussian variables. To ease readability, the
same notation $(\Vi,\Wi)$ is used for the original and the approximated
processes. This should not lead to confusion, as long as the
transition densities are distinguished, as done below.

%s2.3 #&#
\subsection{Property of the observation model}\label{sec:HMM}
The observation model is a degenerate HMM. Let us recall the definition
proposed by \citet{cappe05}. A HMM with not countable state space is
defined as a bivariate Markov chain $(X_i, Y_i)$ with only partial
observations $Y_i$, whose transition kernel has a special structure:
both the joint process $(X_i, Y_i)$ \textit{and} the marginal hidden chain
$(X_i)$ are Markovian.

In our model, $(\Wi)$ is not Markovian, only $(\Vi, \Wi)$ is
Markovian. So set $X_i = (\Vi, \Wi)$, with Markov kernel $R(X_{i-1},
dX_{i})=\pDelta(d\Vi, d\Wi|\Vim, \Wim)$, the transition density of
model (\ref{eq:discretize_ML_2}), and $Y_i = X_i^{(1)}$, the first
coordinate of $X_i$ with transition kernel $F(X, dY) =
\mathbh{1}_{\{ Y=X^{(1)} \}}$. Here, $\mathbh{1}_{x}$ is
the Dirac measure in $x$. Thus, the kernel $F$ is zero almost
everywhere and the HMM is degenerate. This leads to an intrinsic
degeneracy of the particle filter used in the standard HMM toolbox, as
explained below.

Therefore, we consider the observation model as a bivariate Markov
chain $( \Vi, \Wi)$ with only partial observations $\Vi$ whose hidden
coordinate $\Wi$ is \textit{not} Markovian. It is not a HMM but a general
dynamic model as considered by \citet{Andrieu2001}. The hidden process
${\Wi}$ is distributed as
\[
\Wz\sim\mu(d\Wz),\qquad \Wi| (\Wtoim, \Vtoim) \sim K(d\Wi|\Wtoim, \Vtoim)
\]
for some conditional distribution function $K$ and the observed process
$\Vi$ is distributed as
\[
\Vi| (\Wtoi, \Vtoim) \sim G(d\Vi|\Wtoi, \Vtoim)
\]
for some distribution function $G$.
Given the Markovian structure of the pair $(\Vi, \Wi)$, we have
$K(d\Wi
|\Wtoim, \Vtoim)=K(d\Wi|\Wim, \Vim)$ and $G(d\Vi| \Wtoi,\allowbreak
\Vtoim) =
G(d\Vi|\W_{i-1:i}, \Vim)$. To simplify, we use the same notation for
random variables and their realizations and assume that $G(d\Vi|\Wtoi$, $\Vtoim) =G(\Vi|\Wtoi, \Vtoim)\,d\Vi$.
%Let us first write the likelihood
%function of the ideal case where the second coordinate $(\Wt)$ is also
%discretely observed. Let $\Wton$ denote a realization of $(\Wt)$ at
%observation times $t_0,\ldots,t_n$. Since the vector $(\Vi, \Wi)$ is
%Markovian, the likelihood $p(\Vton, \Wton;\theta)$ can be written as
%a product of conditional densities
%$$p(\Vton, \Wton;\theta) = p(\Vz,\Wz;\theta)\prod_{i=1}^n p(\Vi, \Wi|
%where $p(\Vi, \Wi|\Vim, \Wim;\theta)$ is the transition density of $(
%Unfortunately, the density $p(\Vi, \Wi|\Vim, \Wim;\theta)$ has no
%explicit
%form because the diffusion is highly non-linear. Therefore, even when
%$(\Wt)$ is discretely observed at the same time
%points as $(\Vt)$, the likelihood is not
%explicit and the maximum likelihood estimator is not available. In
%this ideal case, minimum contrast estimators
%based on the Euler--Maruyama approximation of the diffusion have been
%proposed by \citet{kessler97}.
%
%When the second coordinate $(\Wt)$ is not observed, the estimation is
%much more difficult. Indeed, although $(\Vt, \Wt)$ is Markovian, the
%single coordinate $(\Vt)$ is not Markovian.
%The process $(\Wt)$ is a latent or hidden variable which has to be
%integrated out to compute the likelihood
%p(\Vton;\theta) = \int\ldots\int p(\Vz,\Wz;\theta)\prod_{i=1}^n p(\Vi,
%Again, the transition density $p(\Vi, \Wi|\Vim, \Wim;\theta)$ is
%generally not available and has to be approximated. We therefore
%introduce an approximate diffusion based on the Euler--Maruyama scheme.

%s2.4 #&#
\subsection{Likelihood function}
We want to estimate the parameter $\theta$ by maximum likelihood of the
approximate model, with likelihood
%
%e3 #&#
\begin{equation}
\label{eq:likelihood_Euler} \pDelta(\Vton;\theta) = \int p(\Vz,\Wz;\theta)\prod
_{i=1}^n \pDelta(\Vi, \Wi|\Vim, \Wim;\theta)\,d
\Wton.
\end{equation}
%
%where $\pDelta(\Vi, \Wi| \Vim, \Wim;\theta) $ is the transition
%density of model (\ref{eq:discretize_ML_2}):
% \Delta} \gamma}\exp\left(-\frac{(\Vi-\Vim-\Delta
% f(\Vim,\Wim))^2}{2\Delta\gamma^2}\right)\\
%&\times& \frac{1}{\sqrt{2\pi\Delta} \sigma(\Vim,
% \Wim)}\exp\left(-\frac{(\Wi-\Wim-\Delta
% b(\Vim,\Wim))^2}{2\Delta\sigma^2(\Vim,
% \Wim)}\right).
It corresponds to a pseudo-likelihood for the exact diffusion.
The multiple integrals of equation (\ref{eq:likelihood_Euler}) are
difficult to handle and it is not possible to maximize the
pseudo-likelihood directly.

A solution is to consider the statistical model as an incomplete data
model. The observable vector $\Vton$ is then part of a so-called
complete vector $(\Vton, \Wton)$, where $\Wton$ has to be
imputed. To maximize the likelihood of the complete data vector $(\Vton,
\Wton)$, we propose to use a stochastic version of the
EM algorithm, namely, the SAEM algorithm
[\citet{delyon99}].
Simulation under the smoothing distribution
$\pDelta(\Wton|\Vton;
\theta)$ is likely to be difficult, and direct simulation of the
nonobserved data
$(\Wton)$ is not possible. A SMC algorithm,
also known as Particle Filtering, provides a way to approximate
this distribution [\citet{doucet01}]. We have adapted this algorithm
to handle a coupled two-dimensional SDE, that is, the unobserved
coordinate is nonautonomous and non-Markovian.
% Thus, we are not in the more well-behaved situation of a hidden
%Markov model.
Then, we combine the SAEM algorithm
with the SMC algorithm, where the unobserved data are filtered at each
iteration step, to estimate the parameters of model
(\ref{eq:discretize_ML_2}). Details on the filtering are given in Section~\ref{sec:filtering}, and the SAEM algorithm is presented in Section~\ref{sec:SAEM}. To prove the convergence of this new SAEM--SMC
algorithm, a nonasymptotic deviation inequality is required for the SMC
algorithm. Then we derive the convergence of the SAEM--SMC algorithm to
a maximum of the likelihood.

%s3 #&#
\section{Filtering}\label{sec:filtering}
\label{sec:Particle}

%s3.1 #&#
\subsection{The filtering problem and the SMC algorithm}
For any bounded Borel function $f\dvtx
\mathbb{R} \mapsto\mathbb{R}$, we
denote $\pi_{n,\theta} f = \EDelta (f(\Wn)|\Vton;\theta
)$,
the conditional expectation under the exact smoothing distribution
$\pDelta(\Wton|\Vton; \theta)$ of the approximate model.
The aim is to approximate this distribution for a fixed value of
$\theta
$. When
included in the stochastic EM algorithm, this value will be the current
value $\widehat{\theta}_m$ at the given
iteration.
For notational simplicity, $\theta$ is omitted in the rest of this section.

We now argue why the HMM point of view is ill-posed for the filtering
problem. Considering the model as a HMM, $X_i=(\Vi,
\Wi)$ is the\vspace*{-1pt} hidden Markov chain and $Y_i= X_i^{(1)}$. But then the
filtering problem $\pi_{n}f$ is the ratio of
$ \int\mu(d\Wz) R(\Xz, d\Xun) F(\Xz; \Yun) \cdots R(\Xnm, d\Xn
) F(\Xnm; \Yn) f(\Xn)$
and\break  $ \int\mu(d\Wz) R(\Xz, d\Xun) F(\Xz; \Yun) \cdots R(\Xnm,
d\Xn) F(\Xnm; \Yn)$.
Since
$F(\Xnm;\allowbreak \Yn)= \mathbh{1}_{\{ \Yn= \Xnm^{(1)} \}}$ and the state
space is continuous, the denominator is zero almost surely and the
filtering problem is ill-posed.

Now consider the model in a more general framework
with the hidden state $\Wi$ not Markovian,
and introduce for $i=1, \ldots, n$ the kernels $H_i$ from $\mathbb{R}$
into itself by
%
%e4 #&#
\begin{eqnarray}
\label{Hif} H_i f(u) &=& \int K( dz|u, \Vim) G(\Vi|u, \Vim, z
)f(z)
\nonumber
\\[-8pt]
\\[-8pt]
&=& \int p_\Delta(\Vi,z|\Vim, u) f(z)\,dz.
\nonumber
\end{eqnarray}
Then $\pi_{n}$ can be expressed recursively by
%
%e5 #&#
\begin{eqnarray}
\label{pinf} \pi_n f &=& \frac{\pi_{n-1} H_n f}{\pi_{n-1} H_n 1} %\nonumber\\
%&=&
= \frac{\int\mu(U_0) \prod_{i=1}^np_{\Delta} (V_i, U_i |
V_{i-1}, U_{i-1}) f(U_n)\,d\Wton}{\int\mu
(U_0) \prod_{i=1}^np_{\Delta} (V_i, U_i |
V_{i-1}, U_{i-1})\,d\Wton}.
\end{eqnarray}
Note that the denominator of \eqref{pinf} is
$\mu H_1 \cdots H_n 1 = p_{\Delta}(V_{0:n})$, which is different from
0 since its support is the real line. Thus,
the filtering problem is well-posed.

The kernels $H_i$ are extensions of the kernels considered by
\citet{delmoral01} in the context of two-dimensional SDEs with hidden
coordinate $\Wt$ autonomous (and thus Markovian).
We do not extend their particle filter
% to a two-dimensional SDE with non-autonomous $U_t$
since it is based on simulation
of both $\Vi$ and $ \Wi$ with transition kernel
$p_\Delta(\Vi,\Wi|\Vim, \Wim)$. They avoid the degeneracy of the
weights by introducing an instrumental function $\psi$ and the weights
are computed as $\psi(\Vi^{(k)}-\Vi)$. The choice of this instrumental
function may influence the numerical properties of the filter.
Therefore, we adopt the general filter proposed by \citet{Andrieu2001}
for a more general dynamic system, that we recall here.

The SMC
algorithm provides a set of $K$ particles
$(\Wtonk)_{k=1, \dots, K}$ and\vspace*{-2pt} weights $(\Weighttonk)_{k=1, \dots,K}$
approximating the conditional smoothing distribution
$\pDelta(\Wton|\allowbreak\Vton)$ [see \citet{doucet01}].
The SMC method relies on proposal distributions $q(\Wi| \Vi,\Vim,
\Wim)$ to sample what we call particles from these
distributions. We write $\Vtoi= (\V_0, \ldots, \Vi)$ and likewise
for $\Wtoi$.
%To avoid the problem of degeneracy of the particles, a resampling step
%is introduced resulting in the following algorithm:

%al1 #&#
\begin{algo}[({SMC algorithm})] \label{SMCalgo}
$ $

%Approximation of the conditional distribution $p(\Wton| \Vton; \theta)$

\begin{itemize}
\item\emph{At time $i=0$:} $\forall k = 1,\ldots, K$:
\begin{enumerate}
\item sample $\Wz^{(k)}$ from $p(\Wz|\Vz)$,
\item compute and normalize the weights:
\[
w_0 \bigl(\Wzk \bigr) = p\bigl (\Vz,\Wzk \bigr),\qquad \Weightz \bigl(\Wzk \bigr) =
\frac{w_0
(\Wzk ) }{\sum_{k=1}^{K} w_0
(\Wzk ) }.
\]
\end{enumerate}
\item\emph{At time $i=1,\ldots, n$:} $\forall k = 1,\ldots, K$:
\begin{enumerate}
\item sample indices $A_{i-1}^{(k)}\sim r(\cdot|
\Weightim(\Wtoim^{(1)}), \ldots, \Weightim(\Wtoim^{(K)}) )$.  Set
$\Wtoim^{\prime(k)} = \Wtoim^{(A_{i-1}^{(k)})}$,
\item sample $\Wik\sim q  (\cdot| \V_{i-1:i}, \W^{\prime(k)}_{i-1}
)$ and set $\Wtoik= (\Wtoim^{\prime(k)}, \Wik)$,
\item compute and normalize the weights
\begin{eqnarray*}
\Weighti\bigl(\Wtoik\bigr) &=& \frac
{w_i (\Wtoik ) }{\sum_{k=1}^{K} w_i (\Wtoik
)}\qquad \mbox{with }
\\
w_i \bigl(\Wtoik \bigr) &=& \frac{\pDelta (\Vtoi, \Wtoik
)}{\pDelta (\Vtoim,\Wtoim^{\prime(k)}  )q (\Wik|
\V
_{i-1:i},\Wtoim^{\prime(k)} ).}
\end{eqnarray*}
\end{enumerate}
\end{itemize}
\end{algo}

The SMC algorithm provides an empirical measure
$\Psi^K_{n} =\break  \sum_{k=1}^K \Weightn(\Wton^{(k)}) \mathbh{1}_{\Wton
^{(k)}} $ which is an
approximation to the smoothing distribution
$\pDelta(\Wton|\Vton)$. A draw from this distribution can be obtained by
sampling an index $k$ from a multinomial distribution with
% weights?)}
probabilities
$\Weightn(\Wton^{(1)}), \ldots,\Weightn(\Wton^{(K)}) $ and setting
the draw $\Wton$ equal to $\Wton= \Wtonk$.\vspace*{2pt}

The variable $A_{i-1}^{(k)}$ plays an important role to discard the
samples with small weights and multiply those with large weights
[\citet{Gordon1993}]. It generates a number of offspring
$N_{i-1}^{(\ell)}$, $ \ell
=1,\ldots, K$,
such that $\sum_{\ell=1}^K N_{i-1}^{(\ell)} = K$ and
$\mathbb{E}(N_{i-1}^{(\ell)}) = K \Weightim(\Wtoiml) $.
Many schemes for $r$ have been presented in the literature, including
multinomial sampling [\citet{Gordon1993}], residual sampling
[\citet{Liu1998}] or stratified resampling [\citet{Doucet2000}]. They
differ in terms of $\var(N_{i-1}^{(\ell)})$
[see \citet{Doucet2000}]. The key property that we need in order to
prove the deviation inequality is that
$\mathbb{E}(\mathbh{1}_{\{ A_{i-1}^{(k)}=\ell\}}) = \Weightim
(\Wtoiml) $.\vspace*{-2pt}

Since our model is not a HMM, the weights $w_i (\Wtoik )
$ cannot be written in terms of a
Markov transition kernel of the hidden path as is usually done. It
follows that the
proposal $q$, which is crucial to ensure good convergence properties,
has to depend on $\Vi$. The
first classical choice of $q$ is $q(\Wi|\V_{i-1:i},\Wim) =
\pDelta(\Wi| \Vim,\Wim)$, that is, the transition density. In this case,
the weight reduces to $w_i (\Wtoik ) =
\pDelta(\Vi| \Vim, \Wtoik)$. A second choice for the proposal
is $q(\Wi|\V_{i-1:i},\Wim) = \pDelta(\Wi|
\V_{i-1:i},\Wim)$, that is, the conditional distribution. In this case,
the weight
reduces to $w_i (\Wtoik ) = \pDelta(\Vi| \Vim,
\Wtoimk)$. Transition densities and conditional distributions are
detailed in Appendix \ref{append:model}. When the two Brownian motions are
independent, as we assume, the two choices are equivalent.

This SMC algorithm is plugged into the EM algorithm to estimate the
parameters. We thus need nonasymptotic convergence results on the SMC
algorithm to ensure the convergence of the EM algorithm. This is
discussed in the next section.

%s3.2 #&#
\subsection{Deviation inequality}
In the literature, deviation inequalities for\break  SMC algorithms only
appear for HMM.
To our knowledge, this is the first nonasymptotic result proposed for a
SMC applied to a non-Markovian hidden path. %Furthermore, recall that
%our original system is a diffusion. So we also want to prove
%non-asymptotic result concerning the original filtering distribution
%and the approximation obtained with SMC.
The only result of this type with SDEs has been obtained by \citet
{delmoral01}, with autonomous second coordinate. Here, we generalize
their deviation inequality to a nonautonomous hidden path.

%We start by a Lemma which generalizes the result of \citet{delmoral01}
%to the particle filter we propose.
For a bounded Borel function $f$,
denote
%$\pi_{\theta} f = \E\left(f(\Wn)|\Vton;\theta\right)$,
%the conditional expectation under the exact smoothing distribution
%$\p(\Wton|\Vton; \theta)$ of the exact diffusion model, and
$\Psi_{n}^K f = \sum_{k=1}^K f(\Wn^{(k)})
\Weight_{n,\theta}(\Wton^{(k)})$, the conditional expectation of $f$
under the empirical measure $\Psi_{n,\theta}^K$ obtained by the SMC
algorithm for a given value of $\theta$. We have the following:

%The following lemma is an extension of the result of \citet{delmoral01}
%to a particle filter adapted to a non-autonomous equation for the
%second coordinate of the system and in which $\Vton$ is not
%resimulated.

%pr1 #&#
\begin{prop}\label{lemma}
Under assumption \emph{(SMC3)}, for any $\varepsilon>0$, and for any bounded
Borel function $f$ on $\mathbb{R}$, there exist constants $C_1$ and
$C_2$, independent of $\theta$, such that
%
%e6 #&#
\begin{eqnarray}
\label{lemma1} \mathbb{P} \bigl(\bigl\llvert \Psi_{n,\theta}^K f
- \pi_{n,\theta} f\bigr\rrvert \geq \varepsilon \bigr)&\leq& C_1
\exp \biggl(-K \frac{\varepsilon
^2}{C_2 \|f\|
^2} \biggr), %\\
\end{eqnarray}
where $\| f\|$ is the sup-norm of $f$.
\end{prop}
The proof is provided in Appendix \ref{append:Proof}. A similar
result can be obtained with respect to the exact smoothing
distribution of the exact diffusion model, under assumptions on the
number of particles and the step size of the Euler approximation.

%s4 #&#
\section{Estimation method}\label{sec:estimation}

%s4.1 #&#
\subsection{SAEM algorithm}\label{sec:SAEM}

The EM algorithm [\citet{dempster77}] is useful in situations where the
direct maximization of the marginal likelihood $\theta\rightarrow
\pDelta(\Vton;\theta)$ is more difficult than the maximization of
the conditional expectation of the complete likelihood %$\theta
$Q(\theta|\theta')=\EDelta [\log
\pDelta(\Vton,\Wton;\theta)|\Vton;\theta'  ]$,
where $\pDelta(\Vton,\Wton;\theta)$ is the likelihood of the complete
data $(\Vton,\Wton)$ of the approximate model
(\ref{eq:discretize_ML_2}) and the expectation is under the
conditional distribution of $\Wton$ given $\Vton$ with density
$\pDelta(\Wton|\Vton;\theta')$.
The EM algorithm is an iterative procedure:
at the $m$th iteration, given the current value $\hthetamm$, the
E-step is the
evaluation of $Q_{m}(\theta)=Q(\theta\vert\hthetamm)$, while
the M-step updates $\hthetamm$ by maximizing $Q_{m}(\theta)$.
To fulfill convergence conditions of the algorithm, we consider the
particular case of a distribution from an exponential family. More
precisely, we assume the following:
\begin{itemize}[(M1)]
\item[(M1)] The parameter space $\Theta$ is an open subset of
$\mathbb{R}^p$. The complete likelihood $\pDelta(\Vton,\Wton;
\theta)$
belongs to a curved exponential family, that is,
$\log\pDelta(\Vton,\Wton;\theta)= - \psi(\theta) +  \langle
S(\Vton,\Wton),\nu(\theta) \rangle$,
where $\psi$ and $\nu$ are two functions of $\theta$, $S(\Vton,\Wton)$ is
known as the minimal sufficient statistic of the complete model,
taking its value in a subset ${\mathcal{S}}$ of $\mathbb{R}^d$, and $
\langle\cdot,
\cdot \rangle$ is the scalar product on $\mathbb{R}^d$.
\end{itemize}
The approximate Morris--Lecar model (\ref{eq:discretize_ML_2})
satisfies this assumption when the scaling parameters $V_1,
V_2, V_3$ and $V_4$ are known. Details of the sufficient statistic $S$ are
given in Appendix \ref{append:sufficient}.

Under assumption (M1),
the E-step reduces to the computation of\break  $\EDelta
[S(\Vton,\allowbreak\Wton)|\Vton;\hthetamm ]$. When this expectation
has no closed form, \citet{delyon99}
propose the Stochastic Approximation EM algorithm (SAEM), replacing
the E-step by a stochastic approximation of $Q_{m}(\theta)$.
The E-step is then
divided into a simulation step (S-step) of the nonobserved data
$(\Wton^{(m)})$ with the conditional density $\pDelta(\Wton|\Vton;
\hthetamm)$ and a stochastic approximation step (SA-step) of $\EDelta
[S(\Vton,\Wton)|\Vton;\hthetamm ]$ with a sequence of
positive numbers $(a_m)_{m\in\mathbb{N}}$ decreasing to
zero. We write $s_m$ for the
approximation of this expectation.
%$$s_{m} =s_{m-1}+ a_m \left[S(\Vton,\Wton^{(m)})-s_{m-1}\right], $$
%where $(a_m)_{m\in\mathbb{N}}$ is a sequence of positive numbers
%decreasing to
%zero.
At the S-step, the simulation under the smoothing distribution is done
by SMC, as explained in Section~\ref{sec:Particle}. We call this algorithm the SAEM--SMC algorithm.
Iterations of the SAEM--SMC algorithm are written as follows:
%
%al2 #&#
\begin{algo}[(SAEM--SMC algorithm)]
$ $

\begin{itemize}
\item\emph{Iteration $0$:} initialization of $ \hthetaz$ and set
$s_{0}=0$.
%= \EDelta\left[S(\Vton,\Wton)|\Vton; \hthetaz\right]$.
%
\item\emph{Iteration $m\geq1$:}

%The $k$-th iteration of the SAEM algorithm is thus
%
\begin{itemize}[\emph{SA-step:}]
\item[\emph{S-step:}] simulation of the nonobserved data
$(\Wton^{(m)})$ with SMC targeting the
distribution $\pDelta(\Wton|\Vton; \hthetamm)$.
\item[\emph{SA-step:}] update $s_{m-1}$ using the stochastic approximation:
%
%e7 #&#
\begin{equation}
\label{stoch_approx} s_{m} =s_{m-1}+ a_{m-1} \bigl[S\bigl(
\Vton,\Wton^{(m)}\bigr)-s_{m-1} \bigr].
\end{equation}
%
%where $(a_m)_{m\in\mathbb{N}}$ is a sequence of positive numbers
%decreasing to zero.
%
\item[\emph{M-step:}] update of $\hthetam$ by $\hthetam=
\operatorname{\arg\max}_{\theta\in\Theta}  (-\psi(\theta
)+\langle s_{m},\nu(\theta)\rangle )$.
\end{itemize}
\end{itemize}
\end{algo}

%At the S-step, the simulation under the smoothing distribution is done
%by SMC, as explained in Section~\ref{sec:Particle}. We call this algorithm the SAEM--SMC algorithm.
Standard errors of the estimators can be evaluated through
the Fisher information matrix. Details are given in Appendix
\ref{append:fisher}.
An advantage of the SAEM algorithm is the low-level dependence on the
initialization $\hthetaz$, due to the stochastic approximation of the
E-step. The other advantage over a Monte Carlo
EM algorithm is the computational time. Indeed, only one simulation of
the hidden variables $\Wton$ is needed in the simulation step, while an
increasing number of simulated hidden variables is required in a
Monte Carlo EM algorithm.

%s4.2 #&#
\subsection{Convergence of the SAEM--SMC algorithm}\label{sec:CVSAEM}

The SAEM algorithm we propose in this paper is based on an approximate
simulation step performed with an SMC algorithm. We prove that even if
this simulation is not exact, the SAEM algorithm still converges
toward the maximum of the likelihood of the approximated diffusion
(\ref{eq:discretize_ML_2}).
This is true because the SMC algorithm has good convergence properties.

Let us be more precise. We introduce a set of convergence assumptions
which are the classic ones for the SAEM algorithm [\citet{delyon99}]:
\begin{itemize}[(SAEM3)]
\item[(M2)] The functions $\psi(\theta)$ and $\nu(\theta
)$ are
twice continuously differentiable on~$\Theta$.
\item[(M3)] The function $\bar s\dvtx \Theta\longrightarrow
\mathcal{S}$ defined by
$ \bar s (\theta) = \int S(\vsmall, u) \pDelta(u|\vsmall;\theta)\,d\vsmall \,du$
is continuously differentiable on $\Theta$.
\item[(M4)] The function $\ell_\Delta(\theta) = \log
\pDelta
(\vsmall,u,\theta)$ is continuously differentiable on $\Theta$ and
$ \partial_\theta\int\pDelta(\vsmall, u;\theta)\,d\vsmall\, du= \int
\partial_\theta\pDelta(\vsmall, u;\theta)\,d\vsmall\, du$.
\item[(M5)] Define $L\dvtx \mathcal{S}\times\Theta\rightarrow
\mathbb{R}$ by
$ L(s,\theta) = -\psi(\theta)+\langle s,\nu(\theta)\rangle$.
There exists a function $\hat\theta\dvtx \mathcal{S}\rightarrow\Theta$
such that
$\forall\theta\in\Theta, \forall s \in\mathcal{S}, L(s,\hat
\theta
(s))\geq L(s,\theta)$.
\item[(SAEM1)] The positive decreasing sequence of the
stochastic approximation $(a_m)_{m \geq1}$ is such that $\sum_{m}
a_m = \infty$ and $\sum_{m}a^2_m < \infty$.
\item[(SAEM2)] $\ell_{\Delta} \dvtx \Theta\rightarrow\mathbb{R}$
and $\hat\theta\dvtx \mathcal{S}\rightarrow\Theta$ are $d$ times
differentiable, where $d$ is the dimension of $S(\vsmall, u)$.
\item[(SAEM3)] For all $\theta\in\Theta$, $\int\|
S(\vsmall, u)\|^2 \pDelta(u|\vsmall;\theta)\,du< \infty$ and the
function $\Gamma(\theta)=\Cov_\theta(S(\cdot,\Wton))$ is continuous,
where the covariance is under the conditional distribution $\pDelta
(\Wton| \Vton; \theta)$.
% \item[\textbf{(SAEM4)}] For any positive Borel function $f$,
% $\EDelta(f(\Wton^{(m+1)})|\mathcal{F}_m) = \int f(\w)\pDelta(\w|$ $
% where $\{\mathcal{F}_m\}$ is the increasing family of $
%$\Wton^{(2)}, \ldots, \Wton^{(m)}$.
%
\item[(SAEM4)] Let $\{\mathcal{F}_m\}$ be the increasing
family of $\sigma$-algebras generated by the random variables $s_0,
\Wton^{(1)}$, $\Wton^{(2)}, \ldots, \Wton^{(m)}$. For any positive
Borel function $f$,
$\EDelta(f(\Wton^{(m+1)})|\mathcal{F}_m) = \int f(u)\pDelta
(u|\vsmall,\hthetam)\,du$.
\end{itemize}
Assumptions (M1)--(M5) ensure the convergence of the EM
algorithm when the E-step is exact [\citet{delyon99}]. Assumptions
(M1)--(M5) and (SAEM1)--(SAEM4) together with the
additional assumption that $(s_m)_{m\geq0}$ takes its values in a
compact subset of
$\mathcal{S}$ ensure the convergence of the SAEM estimates to a
stationary point of the observed likelihood $\pDelta(\Vton;\theta)$
when the simulation step is exact [\citet{delyon99}].

Here the simulation step is not exact and we have three additional
assumptions on the SMC algorithm to bound the error induced by this
algorithm and prove the convergence of the SAEM--SMC algorithm:
\begin{itemize}[(SMC3)]
\item[(SMC1)] The number of particles $K$ used at
each iteration of the SAEM algorithm varies along the
iteration: there exists a function $g(m) \rightarrow\infty$
when $m \rightarrow\infty$ such that
$K(m) \geq g(m) \log(m)$.
\item[(SMC2)] The function $S$ is bounded uniformly in $u$.
\item[(SMC3)] The functions $\pDelta(\V_i|\W_i, \V_{i-1},
\W_{i-1}; \theta)$ are bounded uniformly in $\theta$.
\end{itemize}

%th1 #&#
\begin{theorem}\label{thcv_saem}
$\!\!\!$Assume that \emph{(M1)--(M5)},
\emph{(SAEM1)--(SAEM3)} and \emph{(SMC1)--(SMC3)} hold. Then, with probability 1,
$\lim_{m\rightarrow\infty}$ $ d( \hthetam, \mathcal{L}) = 0 $, where
$\mathcal{L}=\{ \theta\in\Theta, \partial_\theta\ell_\Delta
(\theta)=0\}$ is the set of stationary points of the
log-likelihood\break
$\ell_\Delta(\theta) = \log\pDelta(\Vton;\theta)$.
\end{theorem}
Theorem~\ref{thcv_saem} is proved in Appendix \ref{append:Proof}. Note
that assumption (SAEM4) is not needed thanks to the conditional
independence of the particles generated by the SMC algorithm, as
detailed in the proof. Similarly, the additional assumption that
$(s_m)_{m\geq0}$ takes its values in a compact subset of
$\mathcal{S}$ is not needed, as it is directly satisfied under
assumption (SMC2).

We deduce that the SAEM algorithm converges to a (local) maximum of
the likelihood under standard additional assumptions (LOC1)--(LOC3)
proposed by \citet{delyon99} on the regularity of the log-likelihood
$\ell_\Delta(\Vton;\theta)$ that we do not recall here.
%
%co1 #&#
\begin{corollary}\label{corollary}
Under the assumptions of Theorem~\ref{thcv_saem} and additional
assumptions \emph{(LOC1)--(LOC3)}, the sequence $\hthetam$ converges with
probability 1 to a (local) maximum of the likelihood $p_\Delta(\Vton;
\theta)$.
\end{corollary}
%
%Proof is given in Appendix \ref{append:Proof}.

The classical assumptions (M1)--(M5) are usually satisfied. Assumption\linebreak[4]
(SAEM1) is easily satisfied by choosing properly the sequence $(a_m)$.
Assumptions (SAEM2) and (SAEM3) depend on the regularity of the model.
They are satisfied for the approximate Morris--Lecar model.

%Assumptions (SMC2) and (SMC3) are satisfied for the approximate
%Morris-Lecar model because the variables $\W$ are bounded between 0
%and 1 and the variables $\V$ are fixed at their observed values. This
%would not have been the case with \cite{delmoral01}'s filter which
%resimulate the variables $\V$ at each iteration.
In practice, the SAEM algorithm is implemented with an
increasing number equal to the iteration number, which satisfies
Assumption (SMC1).
Assumption (SMC2) is satisfied for the approximate
Morris--Lecar model because the variables $\W$ are bounded between 0
and 1 and the variables $\V$ are fixed at their observed values. This
would not have been the case with the filter of \citet{delmoral01}, which
resimulates the variables $\V$ at each iteration. Assumption (SMC3)
is satisfied if we require that $\gamma$ is strictly bounded away from
zero; $\gamma\geq\varepsilon> 0$.

%s4.3 #&#
\subsection{Properties of the approximate diffusion}
The SAEM--SMC algorithm provides a sequence which converges to the set
of stationary points of the log-likelihood $\ell_\Delta(\theta) =
\log
p_\Delta(\Vton;\theta)$. The following result aims at comparing this
likelihood, which corresponds to the Euler approximate model
(\ref{eq:discretize_ML_2}), with the true likelihood
$p(\Vton;\theta)$.
The result is based on the bound of the Euler approximation proved by
\citet{Gobet2008}. Their result holds under the following assumption:
\begin{itemize}[(H1)]
\item[(H1)] Functions $f$, $b$, $\sigma$ are 2 times
differentiable with bounded derivatives with respect to $u$ and $v$ of
all orders up to 2.\vadjust{\goodbreak}
\end{itemize}
%
%$u$ and $v$? It should be added.}
Let us assume we apply the SAEM algorithm on an approximate model
obtained with an Euler scheme of step size $\delta=\Delta/L$. Then we
have the following:
%
%th2 #&#
\begin{theorem}\label{th_euler}
Under assumption \emph{(H1)}, there exists a constant $C$, independent of
$\theta$, such that for any $\theta\in\Theta$ and any vector $\Vton$,
\[
\bigl|p(\Vton; \theta)-p_\delta(\Vton; \theta) \bigr|\leq C \frac
{1}{L}n
\Delta.
\]
\end{theorem}
Proof is given in Appendix \ref{append:Proof}.
Assumption (H1) is a strong assumption, which is sufficient and not
necessary. It does not hold for the Morris--Lecar model.
Different sets of weaker assumptions have been proposed to prove the
convergence of the Euler scheme in the strong sense (expectation of the
absolute error between the exact and approximated process). The proofs
are mainly based on localization arguments; see \citet{2012arXiv1204.6620K} for a review
paper. The convergence of the densities has been less studied, and it is
beyond the scope of this paper.

%s5 #&#
\section{Intracellular recordings from a turtle
motoneuron}\label{sec:application}

\begin{figure}

\includegraphics{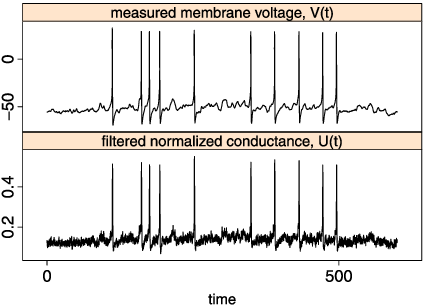}

\caption{Observations of the membrane potential in a spinal
motoneuron of an adult red-eared turtle during 600 ms (upper panel)
and the filtered hidden process of the normalized conductance
associated with K$^+$ current (lower panel) for the estimated
parameters with the scaling parameters fixed at $V_1 = -2.4$ mV,
$V_2 = 36$ mV, $V_3 = 4$ mV and $V_4 = 60$~mV.}\label
{fig:experimentaldatatrace}\vspace*{-3pt}
\end{figure}

The membrane potential from a
spinal motoneuron in segment D10 of an adult red-eared turtle
(\emph{Trachemys scripta elegans}) was recorded while a periodic mechanical
stimulus was applied to selected regions of the carapace with a
sampling step of 0.1 ms [for details see
\citet{Berg2007,BergDitlevsen2008}].
The turtle responds to the stimulus with a reflex
movement of a limb known as the \emph{scratch reflex}, causing an
intense synaptic input to the recorded neuron.
Due to the time-varying stimulus, a model for the complete data set
needs to incorporate the time inhomogeneity, as done in
\citet{Jahnetal2011}. However, in \citet{Jahnetal2011}, only
one-dimensional diffusions are considered, and spikes are
modeled as single points in time by adding a jump term with
state-dependent intensity function to the SDE,
ignoring the detailed dynamics during spikes. In this paper we
aim at estimating parameters during spiking activity by explicit
modeling of time-varying conductances. Therefore, we only
analyze four traces during on-cycles [following
\citet{Jahnetal2011}] where spikes occur. Furthermore, in these time
windows, the input is
approximately constant, which is required for
the Morris--Lecar model with constant parameters. An example of the
analyzed data is plotted in
Figure~\ref{fig:experimentaldatatrace}, together with a filtered trace
of the unobserved coordinate.

%f2 #&#

%t1 #&#
\begin{table}[b]\vspace*{-3pt}
\tabcolsep=0pt
\caption{Parameter estimates obtained from observations of the
membrane potential of a spinal
motoneuron of an adult red-eared turtle during 600 ms for two
different sets of scaling parameters. With $\sigma= 0.05$ fixed.
First trace}\label{tab:realdata1}
\begin{tabular*}{\textwidth}{@{\extracolsep{\fill}}ld{2.3}d{2.3}d{2.3}d{1.3}d{4.3}d{1.3}d{2.3}d{3.3}@{}}
\hline
&\multicolumn{8}{c}{\textbf{Parameter}}\\[-6pt]
&\multicolumn{8}{c}{\hrulefill}\\
& \multicolumn{1}{c}{$\boldsymbol{g_{\mathrm{L}}}$}
& \multicolumn{1}{c}{$\boldsymbol{g_ {\mathrm{Ca}}}$}
& \multicolumn{1}{c}{$\boldsymbol{g_{\mathrm{K}}}$}
& \multicolumn{1}{c}{$\boldsymbol{\gamma}$}
& \multicolumn{1}{c}{$\boldsymbol{V_{\mathrm{K}}}$}
& \multicolumn{1}{c}{$\boldsymbol{\phi}$}
& \multicolumn{1}{c}{$\boldsymbol{V_{\mathrm{Ca}}}$}
& \multicolumn{1}{c}{$\boldsymbol{I}$}\\
\hline
\multicolumn{9}{l}{{With $V_1 = -1.2$ mV, $V_2 = 18$ mV, $V_3 = 2$ mV, $V_4 = 30$ mV}}                                                                          \\
\quad Estimate                                                                               & -0.296 & 11.274 & 6.553  & 2.801 & -124.481 & 1.989 & 35.769 & -5.024  \\
\quad SE                                                                                     & 0.001  & 0.028  & 0.049  & 0.001 & 14.563   & 0.000 & 0.122  & 0.052   \\
[3pt]
\multicolumn{9}{l}{{With $V_1 = -2.4$ mV, $V_2 = 36$ mV, $V_3 = 4$ mV, $V_4 = 60$ mV}}                                                                          \\
\quad Estimate                                                                               & 1.046  & 12.906 & 20.878 & 2.466 & -67.097  & 2.153 & 98.698 & -65.403 \\
\quad SE                                                                                     & 0.009  & 0.008  & 0.021  & 0.001 & 0.227    & 0.001 & 0.198  & 1.204   \\
\hline
\end{tabular*}
\end{table}

First the model was fitted with the values of the scaling parameters
$V_1$--$V_4$ given in \citet{RinzelErmentrout1989} and used in Section~\ref{sec:simulation} below; see Table~\ref{tab:realdata1} for one of the traces. Most of the
estimates are reasonable and in agreement with the expected
order of magnitudes for the parameter values, except for the
$V_{\mathrm{Ca}}$
reversal potential, which in the literature is reported to be around
100--150 mV (estimated to 44.7 mV),\vadjust{\goodbreak} and the leak
conductance, which is estimated to be negative. Conductances are
always nonnegative. This is probably due to wrong choices of the scaling
constants $V_1$--$V_4$. For the parameters given in
\citet{RinzelErmentrout1989}, the average of the membrane potential
$V_t$ between spikes is around $-$26 mV, whereas the average of the
experimental trace between spikes is around $-$56 mV, a factor two
larger. We therefore rerun the estimation procedure fixing
$V_1$--$V_4$ to twice the value from before, which provides
approximately the same values of the normalized Ca$^{2+}$
conductance, $m_{\infty} (\cdot)$, and the rates of opening and
closing of
K$^+$ ion channels, $\alpha(\cdot)$ and $\beta(\cdot)$, as in the
theoretical model when $V_t$
is at its equilibrium value. In this case all parameters are reasonable
and in
agreement with the expected order of magnitudes.\vadjust{\goodbreak}
%In
%Fig.~\ref{fig:SAEMconvergenceRealdata} the convergence of the SAEM
%algorithm is presented. As in the simulated data examples, it is seen
%that the algorithm converges for the parameters of the observed
%coordinate in very few
%iterations to a neighborhood of some value. Only for the parameters of
%the un-observed coordinate, $\phi$ and $\sigma$, more
%iterations are needed.

%For two parameters, $g_L$ and $\sigma$, the
%estimated variances were negative, but very small in absolute
%values. This can be due to numerical instabilities, and should be
%interpreted as being close to zero. The SEs are probably
%underestimated,
%though, as shown in the simulation study.

%t2 #&#
\begin{table}
\def\arraystretch{0.95}
\tabcolsep=0pt
\caption{Parameter estimates obtained from observations of the
membrane potential of a spinal
motoneuron of an adult red-eared turtle during 600 ms for three
different values of $\sigma$. With $V_1 = -2.4$ mV, $V_2 = 36$ mV,
$V_3 = 4$ mV, $V_4 = 60$ mV fixed. First
trace}\label{tab:realdata2}\vspace*{-2pt}
\begin{tabular*}{\textwidth}{@{\extracolsep{\fill}}ld{1.3}d{2.3}d{2.3}d{1.3}d{3.3}d{1.3}d{2.3}d{3.3}@{}}
\hline
%& \multicolumn{9}{c}{Parameters}\\
&\multicolumn{8}{c}{\textbf{Parameter}}\\[-6pt]
&\multicolumn{8}{c}{\hrulefill}\\
& \multicolumn{1}{c}{$\boldsymbol{g_{\mathrm{L}}}$}
& \multicolumn{1}{c}{$\boldsymbol{g_ {\mathrm{Ca}}}$}
& \multicolumn{1}{c}{$\boldsymbol{g_{\mathrm{K}}}$}
& \multicolumn{1}{c}{$\boldsymbol{\gamma}$}
& \multicolumn{1}{c}{$\boldsymbol{V_{\mathrm{K}}}$}
& \multicolumn{1}{c}{$\boldsymbol{\phi}$}
& \multicolumn{1}{c}{$\boldsymbol{V_{\mathrm{Ca}}}$}
& \multicolumn{1}{c}{$\boldsymbol{I}$}\\
\hline
\multicolumn{9}{l}{{$\sigma$ fixed to 0.02}}\\
\quad Estimate &1.302 & 12.460 & 16.550 & 2.280 &-74.301 &2.881 &99.398&
-52.742\\
\quad SE& 0.003& 0.028& 0.131&0.000 &0.703 &0.001 &2.384 &2.727\\
[3pt]
\multicolumn{9}{l}{{$\sigma$ fixed to 0.05} }\\
\quad Estimate &1.046 & 12.906& 20.878 & 2.466 &-67.097 &2.153 &98.698&
-65.403\\
\quad SE&0.009 &0.008 &0.021 &0.001 &0.227 &0.001 &0.198 &1.204\\
[3pt]
\multicolumn{9}{l}{{$\sigma$ fixed to 0.15}}\\
\quad Estimate &1.308 & 12.442 & 16.419 & 2.301 &-74.576 &2.911 &99.417&
-52.341\\
\quad SE& 0.002& 0.007& 0.120&0.000 &0.021 &0.000 &4.529 &1.558\\
\hline
\end{tabular*} \vspace*{-3pt}
\end{table}

%t3 #&#
\begin{table}[b]\vspace*{-3pt}
\def\arraystretch{0.95}
\tabcolsep=0pt
\caption{Parameter estimates obtained from four different traces of the
membrane potential of a spinal
motoneuron of an adult red-eared turtle. Each trace has 6000
observations points with a sampling step of 0.1 ms. With $V_1 =
-2.4$ mV, $V_2 = 36$ mV,
$V_3 = 4$ mV, $V_4 = 60$ mV and $\sigma=0.05$
fixed}\label{tab:realdata3}\vspace*{-2pt}
\begin{tabular*}{\textwidth}{@{\extracolsep{\fill}}ld{1.3}d{2.3}d{2.3}d{1.3}d{3.4}d{1.3}d{3.3}d{3.3}@{}}
\hline
%& \multicolumn{9}{c}{Parameters}\\
&\multicolumn{8}{c}{\textbf{Parameter}}\\[-6pt]
&\multicolumn{8}{c}{\hrulefill}\\
& \multicolumn{1}{c}{$\boldsymbol{g_{\mathrm{L}}}$}
& \multicolumn{1}{c}{$\boldsymbol{g_ {\mathrm{Ca}}}$}
& \multicolumn{1}{c}{$\boldsymbol{g_{\mathrm{K}}}$}
& \multicolumn{1}{c}{$\boldsymbol{\gamma}$}
& \multicolumn{1}{c}{$\boldsymbol{V_{\mathrm{K}}}$}
& \multicolumn{1}{c}{$\boldsymbol{\phi}$}
& \multicolumn{1}{c}{$\boldsymbol{V_{\mathrm{Ca}}}$}
& \multicolumn{1}{c}{$\boldsymbol{I}$}\\
\hline
\multicolumn{9}{l}{{First trace}}\\
\quad Estimate &1.046 & 12.906& 20.878 & 2.466 &-67.097 &2.153 &98.698&
-65.403\\
\quad SE&0.009 &0.008 &0.021 &0.001 &0.227 &0.001 &0.198 &1.204\\
[3pt]
\multicolumn{9}{l}{{Second trace}}\\
\quad Estimate &1.430 & 11.705 & 15.791 & 2.253 &-73.029 &3.138 &103.709&
-53.183\\
\quad SE& 0.003 &0.008 &0.029 &0.000 &0.641 &0.001 &1.269 &0.651\\
[3pt]
\multicolumn{9}{l}{{Third trace}}\\
\quad Estimate &1.371 & 11.878 & 15.379 & 2.210 &-75.024 &3.004 &99.887&
-49.499\\
\quad SE&0.002 &0.013 & 0.017& 0.000& 0.195& 0.000& 0.464&0.614\\
[3pt]
\multicolumn{9}{l}{{Fourth trace}}\\
\quad Estimate &1.197 & 11.452 & 12.521 & 2.012 &-85.982 &3.776 &99.615&
-37.017\\
\quad SE&0.002 &0.055 &0.017 &0.000 &0.089 &0.000 &1.466 &0.861\\
\hline
\end{tabular*}
\end{table}

To check the robustness to misspecifications in the diffusion
parameter $\sigma$ of the unobserved coordinate, we fitted the model
for three different values of $\sigma$; see Table~\ref{tab:realdata2}. Results are stable and suggest that $\sigma$ is
primarily affecting the subthreshold fluctuations of the channel
dynamics, and mainly the spiking dynamics of the unobserved coordinate
influences the first coordinate.

Final results for all four traces are presented in Table~\ref{tab:realdata3}. It is reassuring that the parameter estimates
seem so reproducible over different\vadjust{\goodbreak} traces; the largest variation was
below 10\%. This is not due to starting values, for example, the starting
value for $g_{\mathrm{L}}$ was 0.1, and all four estimates ended up between 1.3
and 1.4, and the starting value for $V_{\mathrm{K}}$ was $-55$, and all
four estimates ended up between $-75.4$ and $-74.4$.

%s6 #&#
\section{Simulation study}\label{sec:simulation}
Parameter values of the Morris--Lecar model used in the simulations are
the same
as those of \citet{RinzelErmentrout1989,Tateno2004} for a class II
membrane, except that we
set the membrane capacitance constant to $C = 1$ $\mu$F$/$cm$^2$, which
is the standard value reported in the literature. Conductances and
input current were correspondingly changed and, thus, the two models
are the same. The values are as follows:
$V_{\mathrm{K}} = -84$ mV, $V_{\mathrm{L}} = -60$~mV, $V_{\mathrm{Ca}} = 120$~mV, $C = 1$
$\mu$F$/$cm$^2$, $g_{\mathrm{L}} = 0.1$ $\mu$S$/$cm$^2$, $g_{\mathrm{Ca}} = 0.22$~$\mu$S$/$cm$^2$, $g_{\mathrm{K}} = 0.4$ $\mu$S$/$cm$^2$, $V_1 = -1.2$ mV, $V_2 = 18$ mV,
$V_3 = 2$ mV, $V_4 = 30$ mV, $\phi= 0.04$~ms$^{-1}$.
Input is chosen to be $I = 4.5$ $\mu$A$/$cm$^2$. Initial conditions of the
Morris--Lecar model are $V_{t_0} = -26$ mV, $\W_{t_0} = 0.2$.
The volatility parameters are $\gamma= 1$~mV~ms$^{-1/2}$,
$\sigma= 0.03$ $\mbox{ms}^{-1/2}$.
Trajectories are simulated with time step $\delta=0.01$ ms and $n=2000$
points are subsampled with observations time step $\Delta= 10 \delta$.
Then $\theta$ is estimated on each simulated trajectory.
A hundred repetitions are used to evaluate the performance of the estimators.
An example of a simulated trajectory (for $n=10\mbox{,}000$) is given in Figure~\ref{fig:simu_ML}.

%f3 #&#
\begin{figure}

\includegraphics{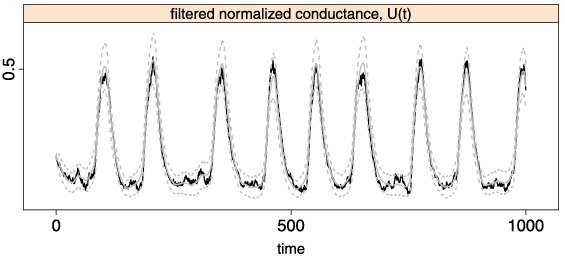}

\caption{Filtering of $(\W_t)$ with the particle filter algorithm (100
particles):
hidden simulated trajectory of the Morris--Lecar model
$(\W_t)$ (black), mean filtered signal (grey full drawn line), 95\%
confidence interval of filtered signal (grey dashed lines).%A: With
% resampling step. B: Without resampling step. C: PIMH algorithm (100
% iterations, 100 particles).
}\label{fig:simu_filter}
\end{figure}

%s6.1 #&#
\subsection{Filtering results}

The Particle filter aims at filtering the hidden process $(\Wt)$ from
the observed process $(\Vt)$. We illustrate its performance on a
simulated trajectory, with $\theta$ fixed at its true value.
The SMC Particle filter algorithm is implemented with $K=100$ particles
and the
transition density as proposal; see
Figure~\ref{fig:simu_filter}. The true hidden process, the mean
filtered signal and
its 95\% confidence interval are plotted. The filtered process appears
satisfactory. The
confidence interval includes the true hidden process $(\Wt)$.

%s6.2 #&#
\subsection{Estimation results}
The performance of the SAEM--SMC algorithm is illustrated on 100
simulated trajectories. The SAEM algorithm is implemented
with $m=200$ iterations and a sequence ($a_m$) equal to 1 during the 100
first iterations and equal to $a_m = 1/(m-100)^{0.8}$ for $m>
100$. The SMC algorithm is implemented with $K(m)=\min(m,100)$ particles
at each
iteration of the SAEM algorithm. The
SAEM algorithm is initialized by a random draw of
$\widehat{\theta}_0$ not centered around the true value:
$\widehat{\theta}_0 = \theta_{\mathrm{true}} + 0.1 + \theta_{\mathrm{true}}/3
\mathcal{N}(0,1)$.

\begin{figure}

\includegraphics{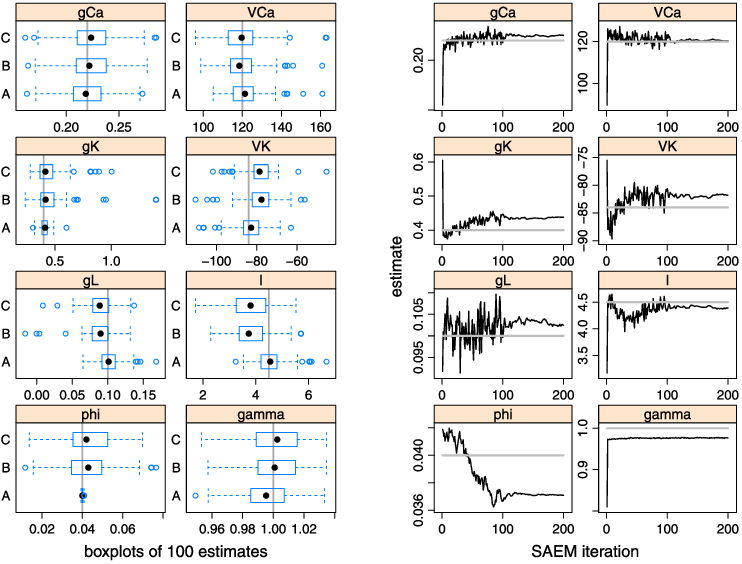}

\caption{Estimation results for simulated data. Left panels:
Boxplots of 100 estimates from simulated data sets for the 8
parameters. True values used in
the simulations are given by
the gray lines. A:~Both $\V_t$ and $\W_t$ are observed. B: Only
$\V_t$ is observed, $\sigma$ is fixed at the true value 0.03. C: Only
$\V_t$ is observed, $\sigma$ is fixed at a wrong value 0.04.
Right panels: Convergence of the SAEM algorithm for the 8 estimated
parameters on a
simulated data set. True values used in the simulation are given by
the gray lines.}\label{fig:boxplots}
\end{figure}

An example of the convergence of the SAEM algorithm for one of the
iterations is presented in Figure~\ref{fig:boxplots}. It is seen
that the algorithm converges for most of the parameters in a few
iterations to a neighborhood of the true value, even if the initial
values are far from the true ones. Only for $\phi$ more
iterations are needed, which is expected since this parameter
appears in the second, nonobserved coordinate.

%f4 #&#

%t4 #&#
\begin{table}
\tabcolsep=0pt
\caption{Simulation results obtained from 100 simulated Morris--Lecar
trajectories ($n=2000$, $\Delta=0.1$ ms). Two estimators are
compared: The pseudo maximum likelihood estimator in the ideal case
where both $\V_t$ and $\W_t$ are observed, and the SAEM estimator
when only $\Vt$ is observed with the SAEM initialization at a
random value not centered around the true value $\theta$. An example
of standard errors (SE) estimated with the SAEM--SMC algorithm on
one single simulated data set is also given}\label{tab:simuSAEM}
\begin{tabular*}{\textwidth}{@{\extracolsep{4in minus 4in}}ld{1.3}d{1.3}d{1.3}d{1.3}d{3.3}d{1.3}d{3.3}d{1.3}@{}}
\hline
&\multicolumn{8}{c@{}}{\textbf{Parameter}}\\[-6pt]
&\multicolumn{8}{c@{}}{\hrulefill}\\
\multicolumn{1}{l}{\textbf{Estimator}}
& \multicolumn{1}{c}{$\boldsymbol{g_{\mathrm{L}}}$}
& \multicolumn{1}{c}{$\boldsymbol{g_ {\mathrm{Ca}}}$}
& \multicolumn{1}{c}{$\boldsymbol{g_{\mathrm{K}}}$}
& \multicolumn{1}{c}{$\boldsymbol{\gamma}$}
& \multicolumn{1}{c}{$\boldsymbol{V_{\mathrm{K}}}$}
& \multicolumn{1}{c}{$\boldsymbol{\phi}$}
& \multicolumn{1}{c}{$\boldsymbol{V_{\mathrm{Ca}}}$}
& \multicolumn{1}{c@{}}{$\boldsymbol{I}$}\\
\multicolumn{1}{l}{\textbf{True values:}} & \multicolumn{1}{c}{$\boldsymbol{0.100}$}
& \multicolumn{1}{c}{$\boldsymbol{0.220}$}
& \multicolumn{1}{c}{$\boldsymbol{0.400}$}
& \multicolumn{1}{c}{$\boldsymbol{1.00}$}
& \multicolumn{1}{c}{$\boldsymbol{-84.00}$}
& \multicolumn{1}{c}{$\boldsymbol{0.040}$} &
\multicolumn{1}{c}{$\boldsymbol{120.00}$} &
\multicolumn{1}{c@{}}{$\boldsymbol{4.400}$}\\
\hline
\multicolumn{9}{l}{With both $\V_t$ and $\W_t$ observed (pseudo maximum likelihood estimator)}\\
\quad Mean & 0.101 & 0.219& 0.411& 0.996 &-83.20& 0.040& 121.97& 4.539\\
\quad RMSE & 0.017 & 0.019& 0.041& 0.019 & 7.61& 0.001 & 8.50& 0.560\\
[3pt]
\multicolumn{9}{l}{With only $\V_t$ observed (SAEM estimator)}\\
\quad Mean & 0.090& 0.225 &0.464 &1.003 &-78.622 &0.041 &119.677 &4.060\\
\quad RMSE & 0.021 & 0.024 &0.144 &0.017 & 9.459 &0.013 &10.218 &1.028\\
[6pt]
\quad Estimated  SE& 0.016 & 0.019 & 0.042 & 0.016 & 4.96 & 0.001 & 7.31 & 0.561\\
\hline
\end{tabular*}
\end{table}

% \begin{figure}
% \centering
% \includegraphics[width = 12cm]{BoxplotsEstimates.pdf}
% \caption{\label{fig:boxplots}
% Boxplots of 100 estimates from simulated data sets for the 9
% parameters. True values used in
% the simulations are given by
% the gray lines. A: Both $\V_t$ and $\W_t$ are observed. B: Only
% $\V_t$ is observed, $\phi$ is fixed at the true value. C: Only
% $\V_t$ is observed, $\phi$ is also estimated.
% }
% \end{figure}

The SAEM estimator is compared with the pseudo maximum likelihood
estimator obtained if both $\V_t$ and $\W_t$ were
observed.
Results are given in Table~\ref{tab:simuSAEM}. The parameters are well
estimated in this ideal case.
The estimation of $\phi$, which is the only parameter in
the drift of the hidden coordinate $\W_t$, is good and does not
deteriorate the estimation of the other parameters. In
Figure~\ref{fig:boxplots} we show boxplots of the estimates of the eight
parameters for the three estimation settings; both coordinates
observed or only one observed with $\sigma$ fixed at either the true
or a wrong value. All parameters appear well estimated.
As expected, the variance of the estimator of $\phi$
hugely
increases when only one coordinate is observed, but interestingly, the
variance of the parameters of the observed coordinate do not seem much
affected by this loss of information.

The SAEM--SMC algorithm provides estimates of the standard errors (SE)
of the estimators (see Appendix \ref{append:fisher}). These should
be close to the RMSE obtained from the 100 simulated data sets. As an
example, the SE for one data set estimated by SAEM are reported in the
last line of Table~\ref{tab:simuSAEM}.
The estimated SE are satisfactory for most
of the parameters, but tend to underestimate.

%s7 #&#
\section{Discussion}

The main contributions of this paper are an algorithm to handle a more
general model than a HMM and to show nonasymptotic convergence results
for the method. It turns out that some of the common
problems encountered with particle filters are not present in our case,
namely, the filter does not degenerate, and we run the algorithm on
large data sets of 6000 observations points in reasonable time (35
minutes on a standard portable computer for one of the simulated data
sets).

To the authors' knowledge, this is the first time
the rate parameter of the unobserved coordinate, $\phi$, is estimated from
experimental data. It is comforting to observe that the estimated
value does not seem to be very sensitive to the choice of scaling
parameters. Other parameters, like the conductances and the reversal
potentials, are more sensitive to this choice, and should be
interpreted with care.

The estimation procedure builds on the pseudo likelihood, which
approximates the true likelihood by an Euler scheme. This
approximation is only valid for a small sampling step, that is, for high
frequency data, which is the case for the type of neuronal data
considered here. If data were sampled less often, a possibility could
be to simulate diffusion bridges between the observed points and
apply the estimation procedure to an augmented data set consisting of the
observed data and the imputed values.

There are several issues that deserve further study. First, it is
important to understand the influence of the scaling parameters
$V_1$--$V_4$ and
how to estimate them for a given data set. The model is not
exponential in these parameters [assumption (M1)] and new estimation
procedures have to be considered. Second, one should be
aware of the possible misspecification of the model. More detailed
models incorporating further types of ion
channels could be explored, but increasing the model complexity might
deteriorate the estimates, since the information contained in only
observing the membrane potential is limited. Furthermore, the sensitivity
on the choice of tuning parameters of the algorithm, like the
decreasing sequence of the stochastic approximation, $(a_m)$, and the number
of SAEM iterations, needs further investigation. Finally, an
automated procedure to find
starting values for the procedure is warranted.

%sA #&#
\begin{appendix}\label{app}
%sB #&#
\section{Distributions of approximate model}\label{append:model}
Consider the general approximate model [see (\ref{eq:discretize_ML_2})]
\[
\pmatrix{ \Vip
\cr
\Wip } %
= \pmatrix{ \Vi
\cr
\Wi }
+ \Delta \pmatrix{ f(\Vi,\Wi)
\cr
b(\Vi,\Wi) } %
+ \sqrt{
\Delta} %
\pmatrix{ \gamma& \rho
\cr
\rho& \sigma(\Vi,\Wi) } %
\pmatrix{ \tilde{\eta}_i
\cr
\eta_i }
\]
with $\rho$ the correlation coefficient between the two Brownian
motions or perturbations.
The distribution of $(\Vip, \Wip)$ conditionally on $(\Vi, \Wi)$ is
\begin{eqnarray*}
&&\hspace*{-3pt}\pmatrix{ \Vip
\cr
\Wip } %
\Big| \pmatrix{ %
\Vi
\cr
\Wi } %
\\
&&\hspace*{-3pt}\qquad\sim%\\
\mathcal{N}%&
\lleft( %
\biggl[
\matrix{ \Vi+ \Delta f(\Vi,\Wi)
\cr
\Wi+ \Delta b(\Vi,\Wi) } %
\biggr],
\Delta \lleft[ %
\matrix{ \bigl(\gamma^2+
\rho^2\bigr) & \rho\bigl(\gamma+\sigma(\Vi,\Wi)\bigr)\vspace*{2pt}
\cr
\rho\bigl(
\gamma+\sigma(\Vi,\Wi)\bigr) & \bigl(\sigma^2(\Vi,\Wi) +
\rho^2\bigr) } %
\rright] \rright).
\end{eqnarray*}

The marginal distributions of $\Vip$ conditionally on $(\Vi, \Wi)$ and
$\Wip$ conditionally on $(\Vi, \Wi)$ are
%
%eB.1 #&#
\begin{eqnarray}
\label{MarginaldistV} \Vip| \Vi, \Wi&\sim& \mathcal{N} \bigl(\Vi+ \Delta f(\Vi,\Wi),
\Delta\bigl(\gamma^2+\rho^2\bigr) \bigr),
\nonumber
\\[-8pt]
\\[-8pt]
\Wip| \Vi, \Wi&\sim& \mathcal{N} \bigl(\Wi+ \Delta b(\Vi,\Wi), \Delta \bigl(
\sigma^2(\Vi,\Wi) + \rho^2\bigr) \bigr).
\nonumber
\end{eqnarray}

The conditional distributions of $\Vip$ conditionally on $(\Wip, \Vi,
\Wi)$ and $\Wip$ conditionally on $(\Vip, \Vi, \Wi)$ are
%
%eB.2 #&#
\begin{eqnarray}
\label{ConditionaldistV} \Vip|\Wip, \Vi, \Wi&\sim& \mathcal{N} (m_V,
\mathit{Var}_V ),
\nonumber
\\[-8pt]
\\[-8pt]
\Wip| \Vip, \Vi, \Wi&\sim& \mathcal{N} (m_U, \mathit{Var}_U
),
\nonumber
\end{eqnarray}
where
\begin{eqnarray*}
m_V &=& \Vi+ \Delta f(\Vi,\Wi) + \frac{\rho(\gamma+\sigma(\Vi,
\Wi))}{
\sigma^2(\Vi,\Wi) + \rho^2}\bigl(\Wip- \Wi-
\Delta b(\Vi,\Wi)\bigr),
\\
\mathit{Var}_V &=& \Delta\bigl(\gamma^2+
\rho^2\bigr) - \frac{\Delta\rho^2(\gamma
+\sigma
(\Vi,\Wi))^2}{ \sigma^2(\Vi,\Wi) + \rho^2},
\\
m_U &=& \Wi+ \Delta b(\Vi,\Wi) +
\frac{\rho(\gamma+\sigma(\Vi,
\Wi))}{
\gamma^2 + \rho^2}\bigl(\Vip- \Vi-\Delta f(\Vi,\Wi)\bigr),
\\
\mathit{Var}_U &=& \Delta\bigl(\sigma^2(\Vi,\Wi) +
\rho^2\bigr) - \frac{\Delta\rho
^2(\gamma+\sigma(\Vi,\Wi))^2}{\gamma^2 + \rho^2}.
\end{eqnarray*}
The distributions in \eqref{MarginaldistV} and
\eqref{ConditionaldistV} are equal when the
Brownian motions are independent, that is, when $\rho=0$.

%%%%%%% Appendix B
%sC #&#
\section{Sufficient statistics}\label{append:sufficient}
%We detail some sufficient statistics functions.
We here provide the sufficient statistics of the approximate model
(\ref
{eq:discretize_ML_2}). Consider the $n\times6$-matrix
\begin{eqnarray*}
X &=& \bigl( - V_{0:(n-1)}, - m_{\infty}(V_{0:(n-1)})
V_{0:(n-1)},
\\
&&\hspace*{5pt} {}- U_{0:(n-1)} V_{0:(n-1)}, U_{0:(n-1)},
\boldsymbol{1}, m_{\infty
}(V_{0:(n-1)})\bigr),
\end{eqnarray*}
where $\boldsymbol{1}$ is the vector of $1$'s of size $n$.
Then the vector
\[
S_1(V_{0:(n-1)},U_{0:(n-1)}) = \bigl(X' X
\bigr)^{-1} X' (V_{1:n}-V_{0:(n-1)})
\]
is the sufficient statistic vector corresponding to the parameters
$\nu_1(\theta) = (g_{\mathrm{L}}, g_{\mathrm{Ca}},\allowbreak g_{\mathrm{K}}, g_{\mathrm{K}} V_{\mathrm{K}}, g_{\mathrm{L}} V_{\mathrm{L}} +I, g_{\mathrm{Ca}}
V_{\mathrm{Ca}})$, where $'$ denotes transposition.

The sufficient statistics corresponding to $\nu_2(\theta) = 1/\gamma
^2$ are
\begin{eqnarray*}
&\displaystyle\sum_{i=1}^n
(\Vi- \Vim )\Wim, \qquad\sum_{i=1}^n
\Wim^2,\qquad \sum_{i=1}^n (
\Vi- \Vim )\Vim m_\infty(\Vim),&
\\
&\displaystyle\sum_{i=1}^n (\Vi- \Vim )
\Wim\Vim,\qquad \sum_{i=1}^n \Wim
^2\Vim ^2.&
\end{eqnarray*}

%$$S_2(V_{0:(n-1)},U_{0:(n-1)})= \frac{1}{n\Delta} \sum_{i=1}^n \left(
%The sufficient statistics corresponding to $\nu_3(\theta) = 1/
%$$\sum_{i=1}^n \left( \alpha(\Vim) (1-\Wim)/\phi-\beta(\Vim) \Wim/\phi
%$$S_3(V_{0:(n-1)},U_{0:(n-1)})= \frac{1}{n\Delta} \sum_{i=1}^n \frac{
The sufficient statistics corresponding to $\phi$ are also explicit but
more complex and not detailed here.

%%%%%%% Appendix C

%sD #&#
\section{Fisher information matrix}\label{append:fisher}

The standard errors (SE) of the parameter estimators can be evaluated
from the diagonal elements of the inverse of the Fisher information
matrix estimate. Its evaluation
is difficult because it has no analytic form. We adapt the estimation
of the Fisher information
matrix, proposed by \citet{delyon99} and based on the Louis' missing
information principle.

The Hessian of the log-likelihood $\ell_\Delta(\theta)$ can be
expressed as
\begin{eqnarray*}
\partial^2_\theta\ell_\Delta(\theta)& =&
\mathbb{E} \bigl[\partial ^2_\theta L\bigl(S(\Vton,\Wton),
\theta\bigr)|\Vton,\theta \bigr]
\\
&&{}+ \mathbb{E} \bigl[\partial_\theta L\bigl(S(\Vton,\Wton), \theta
\bigr) \bigl(\partial _\theta L\bigl(S(\Vton,\Wton), \theta\bigr)
\bigr)'|\Vton,\theta \bigr]
\\
&&{}- \mathbb{E} \bigl[\partial_\theta L\bigl(S(\Vton,\Wton), \theta
\bigr)|\Vton,\theta \bigr] \mathbb{E} \bigl[\partial_\theta L\bigl(S(
\Vton,\Wton), \theta \bigr)|\Vton,\theta \bigr]'.
\end{eqnarray*}
The derivatives $\partial_\theta
L(S(\Vton,\Wton), \theta)$ and
$\partial^2_\theta L(S(\Vton,\Wton), \theta)$ are explicit for the
Euler approximation of the Morris--Lecar model. Therefore, we
implement their estimation using the stochastic approximation procedure
of the SAEM algorithm.
At the $m$th iteration of the algorithm, we evaluate the three
following quantities:
\begin{eqnarray*}
G_{m+1}& =& G_m + a_ {m} \bigl[
\partial_\theta L\bigl(S\bigl(\Vton,\Wton^{(m)}\bigr), \theta
\bigr) - G_m \bigr],
\\
H_{m+1}& =& H_m + a_ {m} \bigl[
\partial^2_\theta L\bigl(S\bigl(\Vton,\Wton ^{(m)}
\bigr), \theta\bigr)
\\
&&\hspace*{46pt} {}+\partial_\theta L\bigl(S\bigl(\Vton,
\Wton^{(m)}\bigr), \theta\bigr) \bigl(\partial _\theta L\bigl(S
\bigl(\Vton,\Wton^{(m)}\bigr), \theta\bigr)\bigr)' -
H_m \bigr],
\\
F_{m+1} &=& H_{m+1}-G_{m+1} (G_{m+1})'.
\end{eqnarray*}
As the sequence $(\hat{\theta}_m)_m$ converges to the maximum of the
likelihood, the sequence $(F_m)_m$ converges to the Fisher information matrix.

%sE #&#
\section{Proof of the convergence results}\label{append:Proof}

%We prove Proposition~\ref{lemma}, Theorem~\ref{thcv_saem}, Corollary~\ref{corollary} and Theorem~\ref{th_euler}.

%%%%%%%%%%%%%%%%%%

%sE.1 #&#
\subsection{Convergence results of Proposition \texorpdfstring{\lowercase{\protect\ref{lemma}}}{1}}

We omit $\theta$ in the proof for clarity.
The conditional expectation $\pi_{n}f$ is
given by (\ref{pinf}) and the kernels $H_i$ from $\mathbb{R}$ into
itself are defined by (\ref{Hif}).
We write $\nu_n = \mu H_1 \cdots H_n 1$ for the constant
conditioned on the observed values $V_{0:n}$. Also, \eqref{Hif} is
bounded, that is, $H_i 1 (u) \leq C$ for all $ u \in\mathbb{R}$ and
$i =
1, \ldots, n$, for some
constant $C$. It directly follows that $\mu H_1 \cdots H_{i-1} 1
\leq C^{i-1}$. Furthermore, we obtain the bound
\[
\mu H_1 \cdots H_{i} 1 \geq\frac{\mu H_1 \cdots H_{i+1} 1}{C} \geq\cdots
\geq\frac{\nu_n}{C^{n-i}}.
\]
Using the above bounds and that $\pi_{i-1}$ is a transition
measure, we obtain
%
%eE.1 #&#
\begin{eqnarray}
\label{bound0} \frac{\nu_n}{C^{n-1}} &\leq& \pi_{i-1}H_i1 \leq
C.
\end{eqnarray}
%
%Consider the SMC sampled particles in algorithm \ref{SMCalgo}.
Define
the two empirical measures obtained at time $i$:
$\Psi^{\prime K}_{i} = \frac{1}{K}\sum_{k=1}^K \mathbh{1}_{\Wtoi
^{\prime(k)}}$ and\vspace*{-1pt}
$\Psi^{K}_{i} = \sum_{k=1}^K W_i(\Wtoi^{(k)})
\mathbh{1}_{\Wtoi^{(k)}}$. We also decompose the weights and write\vspace*{-2pt}
$\Upsilon^K_i f =
\frac{1}{K}\sum_{k=1}^K f(U_i^{(k)})w_i(\Wtoi^{(k)}) $. Then
$W_i(\Wtoi^{(k)}) = w_i(\Wtoi^{(k)})/(K\Upsilon^K_i 1)$ and
$\Psi^{K}_{i} f = \Upsilon^K_i f/\Upsilon^K_i 1$.

Recall the following general result
[\citet{delmoral01}] for $\xi_1, \ldots, \xi_{\mathrm{K}}$ random variables, which
conditioned on a $\sigma$-field $\mathcal{G}$ are independent,
centered and bounded $|\xi_k| \leq a$.
% and $\E(|\xi_i|^2 | \mathcal{G}) \leq b$.
Then for any $\varepsilon> 0$ we have
%
%eE.2 #&#
\begin{eqnarray}
\label{xi} \mathbb{P} \Biggl( \Biggl\llvert \frac{1}{K} \sum
_{k=1}^K \xi_k \Biggr\rrvert \geq
\varepsilon \Biggr) &\leq& 2\exp \biggl( -K\frac{\varepsilon^2}{2a^2} \biggr).
\end{eqnarray}

Let $f$ be a bounded function on $\mathbb{R}$. Then under assumption (SMC3)
\[
\Psi^{\prime K}_{i}f - \Psi^{K}_{i}f
%= \frac{1}{K} \sum_{k=1}^K \left(
% f(U_i^{'(k)}) - K f(U_i^{(k)})W_i(\Wtoi^{(k)}) \right)
= \frac{1}{K} \sum_{k=1}^K
\bigl( f\bigl(U_i^{\prime(k)}\bigr) - \Psi^{K}_{i}f
\bigr) =\frac{1}{K} \sum_{k=1}^K
\xi_k
\]
fulfills the conditions\vspace*{-1pt} for \eqref{xi} to hold with $a=2\| f \|$, since
$\E
( f(U_i^{\prime(k)}) | \mathcal{G}) = \Psi^{K}_{i}f $,
where $\mathcal{G}$ is the $\sigma$-algebra generated by $\Wtoi^{(k)}$.
Thus, for
any $\varepsilon>0$, % we obtain
%
%eE.3 #&#
\begin{eqnarray}
\label{boundb} \mathbb{P} \bigl(\bigl\llvert \Psi^{\prime K}_{i}f
- \Psi^{K}_{i}f\bigr\rrvert \geq \varepsilon \bigr)&\leq& 2
\exp \biggl( -K \frac{\varepsilon
^2}{8\|f\|^2} \biggr).
\end{eqnarray}
Define $Q_i(f)(u) = \int q(u'|V_i, V_{i-1},u) f(u')\,du'$. By
definition of the unnormalized weights in step 3 of the SMC algorithm,
$w_i(u,u')= p_{\Delta} (V_i, V_{i-1},u,u') /$ $ p_{\Delta} (V_{i-1},u)
q(u'|V_i, V_{i-1},u)$, so that $Q_i(fw_i)(u) = \int p_{\Delta} (V_i,u'
|V_{i-1},u)\times\break  f(u')\,du'= H_i f(u) $. We
therefore have
%Likewise, as $H_i f(u) = Q_i(fw_i)(u)$ with $Q_i(f)(u) = \int
%q(u'|V_i, V_{i-1},u) f(u')du'$, we have
\[
\Upsilon^{K}_{i}f - \Psi^{\prime K}_{i-1}H_i
f = \frac{1}{K} \sum_{k=1}^K \bigl(
f\bigl(U_i^{(k)}\bigr)w_i\bigl(
\Wtoi^{(k)}\bigr) -Q_i(fw_i)
\bigl(U_{i-1}^{\prime(k)}\bigr) \bigr) =\frac{1}{K} \sum
_{k=1}^K\xi_k,
\]
which fulfills the conditions for \eqref{xi} to hold, now with $a=2C\| f
\|$ and $\mathcal{G}$ is the $\sigma$-algebra generated by
$\Wtoim^{\prime(k)}$, since $U_i^{(k)}$ is drawn from $q(\cdot|
V_{i-1:i}, U_{i-1}^{\prime(k)})$; see step~2 of the SMC algorithm. Hence,
for any $\varepsilon>0$ we obtain
%
%eE.4 #&#
\begin{eqnarray}
\label{bounda} \mathbb{P} \bigl(\bigl\llvert \Upsilon^{K}_{i}f
- \Psi^{\prime K}_{i-1}H_i f\bigr\rrvert \geq
\varepsilon \bigr)&\leq& 2\exp \biggl( -K \frac{\varepsilon
^2}{8C^2\|f\|^2} \biggr).
\end{eqnarray}
We want to show the following two bounds:
%
%eE.5 #&#
%eE.6 #&#
\begin{eqnarray}
\label{bound1} \mathbb{P} \bigl(\bigl\llvert \Psi^{K}_{i}f
- \pi_{i} f\bigr\rrvert \geq \varepsilon \bigr) &\leq& 2
I_i\exp \biggl( -K \frac{\varepsilon
^2}{8J_i \|f\|^2} \biggr),\qquad i=1, \ldots, n,
\\
\label{bound2} \mathbb{P} \bigl(\bigl\llvert \Psi^{\prime K}_{i}f
- \pi_{i} f\bigr\rrvert \geq \varepsilon \bigr) &\leq& 2
I_i^{\prime}\exp \biggl( -K \frac
{\varepsilon
^2}{8J_i^{\prime} \|f\|^2} \biggr),\qquad
i=0,1, \ldots, n,
\end{eqnarray}
by induction on $i$, for some constants $I_i,I_i^{\prime},J_i,J_i^{\prime}$
increasing with $i$ to
be computed later. Note first that since $\pi_0 = \mu$ and
$U_0^{\prime(k)}$ are i.i.d. with\vspace*{-1pt} law $\mu$, then \eqref{xi} with $\xi_k
= f(U_0^{\prime(k)})-\mu(f)$ yields \eqref{bound2} for $i=0$ with
$I_i^{\prime}= J_i^{\prime}=1$. Let $i\geq1$ and assume \eqref{bound2} holds
for $i-1$. We can write
\[
\Psi_i^K f - \pi_i f = \frac{1}{\pi_{i-1}H_i 1}
\biggl( \frac{\Upsilon_i^K f}{\Upsilon_i^K 1}\bigl(\pi_{i-1}H_1 1 -
\Upsilon_i^K 1\bigr) + \bigl(\Upsilon_i^K
f - \pi_{i-1}H_i f \bigr) \biggr).
\]
Note that $\Upsilon_i^K 1 > 0$ because the weights $w_i$ are strictly
positive. Define $L_if = \Upsilon_i^K f - \pi_{i-1}H_i f$ and use that
$|\Upsilon_i^Kf| \leq\| f \| \Upsilon_i^K 1$ (because $f$ is bounded)
and \eqref{bound0} to
see that
\begin{eqnarray*}
\bigl|\Psi_i^K f - \pi_i f\bigr| &\leq&
\frac{C^{n-1}}{\nu_n}\bigl ( \|f \| |L_i1| + |L_if| \bigr)
\end{eqnarray*}
and
\begin{eqnarray*}
|L_i f| &\leq& \bigl| \Upsilon_i^K f -
\Psi_{i-1}^{\prime K} H_i f\bigr| + \bigl|\Psi_{i-1}^{\prime K}
H_i f - \pi_{i-1}H_i f\bigr|.
\end{eqnarray*}
Assuming that \eqref{bound2} holds for $i-1$ and using \eqref{bounda}
and that $\| H_i
f\|\leq C\| f\|$ yield
\begin{eqnarray*}
\mathbb{P} \bigl(\llvert L_if\rrvert \geq \varepsilon
\bigr)&\leq& 2\exp \biggl( -K \frac{\varepsilon
^2}{32C^2\|f\|^2} \biggr) +2I_{i-1}^{\prime}
\exp \biggl( -K \frac
{\varepsilon
^2}{32J_{i-1}^{\prime}C^2\|f\|^2} \biggr).
\end{eqnarray*}
We obtain
\begin{eqnarray*}
&& \mathbb{P} \bigl( \bigl|\Psi^K_i f -
\pi_i f \bigr| \geq \varepsilon \bigr)
\\
&&\qquad\leq\mathbb{P} \biggl( |L_i1| \geq\frac{ \varepsilon\nu_n}{2C^{n-1}
\| f \| } \biggr)
+ \mathbb{P} \biggl( |L_if| \geq\frac{
\varepsilon\nu_n}{2C^{n-1}
} \biggr)
\\
&&\qquad\leq 4\exp \biggl( -K\frac{\varepsilon
^2\nu_n^2}{128C^{2n}\|f\|^2} \biggr) +4I_{i-1}^{\prime}
\exp \biggl( -K \frac{\varepsilon^2\nu_n^2}{128J_{i-1}^{\prime}C^{2n}\|f\|^2} \biggr).
\end{eqnarray*}
Hence, \eqref{bound1} holds with $I_i\geq2(1+I_{i-1}^{\prime})$ and $J_i
\geq16C^{2n}J_{i-1}^{\prime}/\nu_n^2 \geq16J_{i-1}^{\prime}$\vspace*{1pt} since $\nu_n\leq
C^n$. By
\eqref{boundb} and \eqref{bound1} we then conclude that \eqref
{bound2} also
holds for $i$ if $I_i^{\prime} = 1+I_i$ and $J_i^{\prime}= 4 J_i$. These
conditions are fulfilled by choosing $I_i = 3^{i+1}-3$ and $J_i = 16^i$.
Thus, \eqref{lemma1} holds with $C_1=6(3^n-1)$ and $C_2=8 \cdot
16^n$.
This concludes the proof.

%sE.2 #&#
\subsection{Proof of Theorem \texorpdfstring{\lowercase{\protect\ref{thcv_saem}}}{1}}
To prove the convergence of the SAEM--SMC algorithm, we study the
stochastic approximation scheme used during the SA step. The
scheme (\ref{stoch_approx}) can be decomposed into
\[
s_{m+1} = s_{m} + a_m h(s_m) +
a_m e_m + a_m r_m
\]
with
\begin{eqnarray*}
h(s_m) &=&\pi_{n,\hat\theta(s_m)} S -s_m,
\\
e_m &=&S\bigl(\Vton,\Wton^{(m)}\bigr) -
\Psi_{n,\hat\theta(s_m)}^{K(m)}S,
\\
r_m &=& \Psi_{n,\hat\theta(s_m)}^{K(m)}S- \pi_{n,\hat\theta(s_m)} S,
\end{eqnarray*}
where we denote by $\pi_{n,\theta} S = \EDelta(S(\Vton,\Wton
)|\Vton;
\theta)$ the expectation of the sufficient statistic $S$ under
the exact distribution $\pDelta(\Wton|\Vton;\theta)$, and by
$ \Psi_{n,\hat\theta(s_m)}^{K(m)}S$ the expectation of the sufficient
statistic $S$ under the
empirical measure obtained with
the SMC algorithm with $K(m)$ particles and current value of
parameters $\hat\theta(s_m)$ at iteration $m$ of the SAEM--SMC algorithm.

Following Theorem~2 of \citet{delyon99} on the convergence of the
Robbins--Monro scheme, the convergence of the SAEM--SMC algorithm is
ensured if we prove the following assertions:
\begin{enumerate}
\item The sequence $(s_m)_{m\geq0}$ takes its values in a compact set.
\item The function $V(s)=-\ell_\Delta(\hat\theta(s))$ is such that
for all $s\in\mathcal{S}$, $F(s) = \langle\partial_s
V(s),\allowbreak h(s)\rangle
\leq0$ and such that the set $V(\{s, F(s) = 0\})$ is of zero measure.
\item$\lim_{m\rightarrow\infty} \sum_{\ell=1}^m a_\ell e_\ell$ exists
and is finite with probability 1.
\item$\lim_{m\rightarrow\infty} r_m=0$ with probability 1.
\end{enumerate}

Assertion 1 follows from assumption (SMC2) and by construction of $s_m$
in formula (\ref{stoch_approx}).
Assertion 2 is proved by Lemma~2 of \citet{delyon99} under assumptions
(M1)--(M5) and (SAEM2).
Assertion 3 is proved similarly as Theorem~5 of \citet{delyon99}.
By construction of the SMC algorithm, the equivalent of assumption
(SAEM3) is checked for the expectation taken under the approximate
empirical measure $\Psi_{n;\hthetam}^{K(m)}$. Indeed, the assumption of
independence of the nonobserved variables $\Wton^{(1)},\ldots, \Wton
^{(m)}$ given $\widehat{\theta}_{0}, \ldots, \hthetam$ is verified.
As a consequence, for any positive Borel function $f$,
$\EDelta^{K(m)}(f(\Wton^{(m+1)})|\mathcal{F}_m) = \Psi
^{K(m)}_{n;\hthetam} f $.
Then $\sum_{\ell=1}^m a_\ell e_\ell$ is a martingale, bounded in
$L_2$ under assumptions (M5) and (SAEM1)--(SAEM2).

To verify assertion 4, we use Proposition~\ref{lemma}.
Under assumptions (SMC2)--(SMC3) and assertion 1, Proposition~\ref{lemma}
yields that for any $\varepsilon>0$, there exist two constants $C_1$,
$C_2$, independent of $\theta$, such that
\begin{eqnarray*}
\sum_{m=1}^M \mathbb{P}
\bigl(|r_m|>\varepsilon \bigr) &=& \sum_{m=1}^M
\mathbb{P} \bigl(\bigl\llvert \Psi_{n,\hat{\theta}(s_m)}^{K(m)} S - \pi
_{n,\hat
{\theta}(s_m)} S\bigr\rrvert \geq\varepsilon \bigr)
\\
&\leq& C_1 \sum
_{m=1}^M\exp \biggl(-K(m) \frac{\varepsilon^2}{C_2
\|S\|^2}
\biggr).
\end{eqnarray*}
Finally, assumptions (SMC1)--(SMC2) imply that there exists a constant
$C_3$, independent of $\theta$, such that
\[
\sum_{m=1}^M \mathbb{P}
\bigl(|r_m|>\varepsilon \bigr)\leq C_1 \sum
_{m=1}^M\frac{1}{m^{C_3 g(m) \varepsilon^2}},
\]
which is finite when $M \rightarrow\infty$, proving the
a.s. convergence of $r_m$ to 0.

%%\begin{proof}
%Theorem~6 of \citet{delyon99} can be extended without difficulty to our
%algorithm. It proves that under assumptions of Theorem~\ref{thcv_saem}
%and (LOC1), the sequence $\hthetam$ converges to a fixed point of the
%EM-mapping $T(\hthetam) = \hat\theta(s(\hthetam))$. Assumptions
%(LOC2)-(LOC3), Lemma~3 of \citet{delyon99} and application of
%probability 1 to a proper maximum of the likelihood.
%%\end{proof}

%sE.3 #&#
\subsection{Proof of Theorem \texorpdfstring{\lowercase{\protect\ref{th_euler}}}{2}}
The Markov property yields
\begin{eqnarray*}
&&\bigl|p(\Vton;\theta) - p_\delta(\Vton;\theta)\bigr|\\
&&\qquad\leq\int\bigl\llvert p(
\Vton,\Wton;\theta)- p_\delta(\Vton,\Wton;\theta) \bigr\rrvert \,d\Wton
\\
&&\qquad\leq\int\Biggl\llvert \prod_{i=1}^{n}
p (\Vi,\Wi| \V_{i-1},\W_{i-1};\theta ) - \prod
_{i=1}^{n} p_\delta (\Vi,\Wi|
\V_{i-1},\W_{i-1};\theta ) \Biggr\rrvert \,d\Wton
\\
&&\qquad\leq\int\sum_{i=1}^{n}
\bigl\llvert p (\Vi,\Wi| \V_{i-1},\W _{i-1};\theta )-
p_\delta (\Vi,\Wi| \V_{i-1},\W _{i-1};\theta ) \bigr
\rrvert
\\
&&\hspace*{36pt}\qquad{}\times\prod_{j=1}^{i-1}
p (\V_j,\W_j| \V_{j-1},\W_{j-1};
\theta )
\\
&&\hspace*{36pt}\qquad{}\times \prod_{j=i+1}^{n}p_\delta
(\V_j,\W_j| \V_{j-1},\W _{j-1};
\theta )\,d\Wton. %&\leq&\sum_{i=1}^{n}\int\left| p\left(\Vi,\Wi| \V_{i-1},\W_{i-1};
\end{eqnarray*}
\citet{Gobet2008} provide that under assumption (H1), there exist
constants $C_1>0$, $C_2>0$, $C_{3}>0$, $C_{4}>0$ independent of
$\theta$ such that
\begin{eqnarray*}
&&\bigl|p_\delta(\V_i,\W_i|\V_{i-1},
\W_{i-1}; \theta) +p(\V_i,\W_i|\V
_{i-1},\W _{i-1}; \theta)\bigr|
\\
&&\qquad\leq C_{1} e^{-C_{2}\|(\V_i,\W_i)-(\V_{i-1},\W
_{i-1})\|^2},
\\
&&\bigl|p_\delta(\V_i,\W_i|\V_{i-1},
\W_{i-1}; \theta) - p(\V_i,\W_i|\V
_{i-1},\W _{i-1}; \theta)\bigr|
\\
&&\qquad\leq \delta C_{3} e^{-C_{4}\|(\V_i,\W_i)-(\V
_{i-1},\W
_{i-1})\|^2}.
\end{eqnarray*}
%
%Lemma~2 from \citet{donnet08} yields that there exists a constant
%$C_3>0$ independent of $\theta$ such that
% |\pDelta(\V_i,\W_i|\V_{i-1},\W_{i-1}; \theta) - p(\V_i,\W_i|\V_{i-1},
%Set $\mathcal{D}_i = \{\W_i; \|(\V_i,\W_i)-(\V_{i-1},\W_{i-1})\|>2
We deduce that for all $i=1, \ldots,n$, there exists a constant $C >
0$ independent of $\theta$ such that
\begin{eqnarray*}
&&\int\bigl\llvert p (\Vi,\Wi| \V_{i-1},\W_{i-1};\theta )-
p_\delta (\Vi,\Wi| \V_{i-1},\W_{i-1};\theta ) \bigr
\rrvert
\\
&&\qquad{}\times\prod_{j=1}^{i-1} p (
\V_j,\W_j| \V_{j-1},\W_{j-1};\theta
)
\\
&&\qquad {}\times\prod_{j=i+1}^{n}p_\delta
(\V_j,\W_j| \V_{j-1},\W_{j-1};
\theta )\,d\Wton \leq C \delta.
\end{eqnarray*}
Finally, we get
$|p(\Vton;\theta) - p_\delta(\Vton;\theta)|\leq Cn \delta= C\frac{1}L
n\Delta$.
\end{appendix}

\section*{Acknowledgments}

The authors are grateful to Rune W. Berg for making his experimental data
available. We thank E. Gobet for helpful discussions about convergence
of the Euler scheme. The work is part of the Dynamical Systems
Interdisciplinary Network, University of Copenhagen.

%suskaldyti doi

% imsref loaded by aiste.veprauskaite, 2014-04-08 10:00:07

\printaddresses

\end{document}